\documentclass[11pt]{article}
\usepackage{amssymb,amsmath,graphicx,epsfig,setspace,amsfonts,verbatim,natbib}

\usepackage{array}
\usepackage{rotating,longtable,booktabs, colortbl}       % needed for 
\usepackage{mdwlist}
\usepackage[shortlabels]{enumitem}
\usepackage{pdflscape,afterpage}                    % supplies a landscape
\usepackage{appendix}
\usepackage{url,hyperref}                                            % for emails. Also, it links all your references...

\usepackage[standard]{ntheorem}
\usepackage[algo2e]{algorithm2e}

\theoremstyle{break}
\theorembodyfont{\normalfont}

\begin{document}
	\title{Tuning into Climate Risks: Extracting Innovation from Television 
	News for Clean Energy Firms}
	\author{
		Wasim Ahmad\thanks{Email addresses: {\tt wasimad@iitk.ac.in}, {\tt  
		marshad@iitk.ac.in}, {\tt suruchi@iitk.ac.in}, and {\tt 
		preetir@iitism.ac.in}.} 
	    \\ Indian Institute of Technology Kanpur \and
		Mohammad Arshad Rahman \\ Indian Institute of Technology 
		Kanpur \and
		Suruchi Shrimali \\ Indian Institute of Technology 
		Kanpur \\ \and
		Preeti Roy \\ Indian Institute of Technology 
		Dhanbad
		}
	\date{\today}
	\maketitle

\begin{abstract}
\noindent This article develops multiple novel climate risk measures (or 
variables) based on the television news coverage by Bloomberg, CNBC, and Fox 
Business, and examines how they	affect the systematic and idiosyncratic risks 
of clean energy firms in the United States. The measures are built on 
climate related keywords and cover the volume of coverage, type of coverage 
(climate crisis, renewable energy, and government \& human initiatives), and 
media sentiments. We show that an increase in the 
aggregate measure of climate risk, as indicated by coverage volume, reduces 
idiosyncratic risk while increasing systematic risk. When climate risk is 
segregated, we find that systematic risk is positively 
affected by the \textit{physical risk} of climate crises and 
\textit{transition risk} from government \& human initiatives, but no such 
impact is evident for idiosyncratic risk. Additionally, we observe an 
asymmetry in risk behavior: negative sentiments tend to decrease 
idiosyncratic risk and increase systematic risk, while positive sentiments 
have no significant impact. These findings remain robust to including 
print media and climate policy uncertainty variables, though some deviations 
are noted during the COVID-19 period.
	
\vspace{0.5pc}
\noindent \texttt{Keywords:} Climate change, climate finance, 
GDELT, media sentiment, physical risk, transition risk. 
\end{abstract}

%When we talk about an "extra source of market risk" being priced into 
%financial markets, we're referring to additional risks that investors 
%believe could affect asset prices and which are accounted for in the pricing 
%of securities. 

\newlength{\single}\setlength{\single}{1.0\baselineskip}\newlength{\double} %
\setlength{\double}{1.5\baselineskip}\baselineskip\double

%------------------------------------------------------------------------------
\section{Introduction}\label{sec:Intro}
%------------------------------------------------------------------------------

Climate change has become a pressing reality, prompting governments worldwide 
to address the \textit{physical risk} (damages and losses to property due to 
physical consequences of climate change such as hurricanes) of the climate 
crisis and develop policies to manage \textit{transition risk} (the 
potential costs to society in evolving to a low carbon economy to mitigate 
climate change) \citep{NGFS-2024}. A key component of these policies involves 
reducing carbon emissions from fossil fuel consumption and advancing clean 
energy solutions for a sustainable future i.e., mitigating and adapting to 
climate change \citep{Kreibiehl-etal-2022}. However, clean energy 
companies$-$a vital instrument in climate change mitigation$-$face 
stiff competition for investment in the financial market, which is closely 
tied to their risk profile \citep{BloombergNEF-2024}. According to asset 
pricing models \citep[e.g., Capital Asset Pricing Model,][]{Merton-1973}, the 
total risk is divided into systematic (or market) risk and idiosyncratic (or 
unsystematic) risk. Idiosyncratic (systematic) risk can (cannot) be reduced 
or eliminated through diversification and so (higher risk is associated with 
higher expected returns) does not command a risk premium. However, investor 
irrationality resulting from environmental, social, and governance (ESG) 
criteria may limit portfolio diversification and consequently idiosyncratic 
risks may positively affect expected stock returns  
\citet{Levy-1978, Merton-1987, Goyal-Santa-Clara-2003, Pastor-etal-2021, 
Roy-etal-2022, Jagannathan-etal-2023}. Research on the factors influencing 
systematic and/or idiosyncratic risks is expanding, particularly in climate 
finance. Amongst the factors, two determinants that have garnered attention 
are climate risk measures and sentiment variables constructed from print 
media sources. However, the development of these measures from television 
news coverage and subsequent econometric analysis remains unexplored. 
This paper addresses the gap by developing climate risk metrics and sentiment 
variables using snippets from three television news channels--Bloomberg, 
CNBC, and Fox Business--and assessing their impact on both idiosyncratic and 
systematic risks for clean energy firms in the United States (US).

At a broader level, this paper advances the literature on utilizing media 
news data--encompassing textual, visual, or audio elements--to deepen our 
understanding of the financial market, particularly concerning climate risk 
and sentiment metrics. Enhanced media coverage raises public awareness 
about climate change \citep{Sampei-Ayogi-Usui-2009}, leading to more informed 
investment decisions and improved returns on sustainability stock indices 
\citep{El-Ouadghiri-etal-2021}. Within the realm of climate risk research 
(and sentiment analysis, discussed later), the focus has been on extracting 
information from textual sources in print media, 
such as newspapers and financial magazines. For instance, 
\citet{Engle-etal-2020} analyze textual content in \textit{The Wall Street 
Journal (WSJ)} to develop an aggregate measure of climate risk and propose a 
dynamic strategy for hedging climate risk. \citet{Faccini-etal-2023} utilize 
texts from \textit{Reuters} climate-change news to segregate different types 
of climate risks and find that only transition risk from government 
intervention is priced in the US stocks. \citet{Ardia-etal-2023} use content 
from US newspapers and newswires to demonstrate that green stocks tend to 
outperform brown stocks when climate change concerns rise. In a similar vein, 
\citet{Bessec-Fouquau-2024} analyze newspaper contents and find that excess 
stock returns of green and brown stocks are sensitive to newspaper coverage 
of environmental issues. Additionally, 
\citet{Venturini-2022} reviews different types of data required for modeling 
subdivisions of physical and transition risks, and how they affect the 
cross-section of stock returns. 
Other articles, a brief and incomplete list 
we must mention, that make use of textual sources to construct measures of 
physical and transition risks (to examine various issues) include 
\citet{Stroebel-Wurgler-2021}, \citet{Bua-etal-2024}, \citet{Li-etal-2024}, 
and \citet{Kolbel-etal-2024}.

We deviate from existing literature and instead construct (for the first 
time) several measures of climate-related risks derived from television news 
coverage of climate change by Bloomberg, CNBC, and Fox Business. Drawing 
inspiration from \citet{Engle-etal-2020}, 
we create a composite measure of climate risk based on the frequency of 
climate-change snippets aired each month, with each snippet being a 15-second 
block of coverage excluding advertisements. This aggregate measure 
essentially assesses the extent of coverage related to climate change. 
Building on the work of \citet{Faccini-etal-2023} and others, we further 
analyze this coverage by categorizing climate risk into three themes: 
\textit{climate crisis}, \textit{renewable energy}, and \textit{government 
\& human initiative}. The first theme pertains to physical risks, while the 
latter two are associated with transition risks which are particularly 
relevant to market participants and financial news outlets. This 
classification is important because investors react differently to various 
types of climate risks and attune their investment accordingly. 
We evaluate both the overall and thematic climate risks, providing a thorough 
analysis of their effects on the systematic and idiosyncratic risks faced by 
clean energy firms.

% Fourth Paragraph
%-----------------------------------------------------------------------------

Apart from climate risks, researchers have paid great 
attention to extracting media sentiments from textual sources and 
incorporating them into asset price 
models. \citet{TetlockSent-2007} pioneered this approach by developing 
a measure of media pessimism from the content of ``Abreast of the Market'' 
column from the \textit{WSJ}, and showed that media pessimism had significant 
explanatory power for predicting stock returns. Expanding on this work,   
\citet{Tetlock-etal-2008} utilize negative words from articles published in 
\textit{WSJ} and \textit{Dow Jones News Service (DJNS)} about individual S\&P 
500 companies and find that pessimism impact both stock returns and future 
cash flows. \citet{Dougal-etal-2012} also report a similar finding. Taking 
the analysis further, \citet{Garcia-2013} examined two \textit{New York 
Times} columns spanning over a century (1905–2005) by analyzing the frequency 
of positive and negative words in the text; and found that language tone is 
correlated with future stock returns, particularly during recessions. 
\citet{Huang-etal-2014} report an asymmetric effect, where investors react 
more strongly to negative texts than to positive ones. A rich source on
textual sentiment literature is the review article by 
\citet{Kearney-Liu-2014} and references therein. More recent works that find 
evidence of asymmetric media sentiments include \citet{Heston-Sinha-2017}, 
\citet{Bajo-Raimondo-2017}, \citet{Huang-etal-2018}, \citet{Jia-etal-2023}, 
and \citet{He-etal-2024}. Other articles in this genre, to name a few, 
include \citet{Engle-etal-2020}, and \citet{Bask-etal-2024}.

Our research distinguishes itself from previous studies by 
utilizing television news channels as our data source, in contrast to print 
media. We construct two sentiment indices--positive and negative--using the 
NRC Emotion Lexicon (EmoLex), which encompasses two sentiments and eight 
emotions \citep{Mohammad-Turney-2012}. The positive (negative) 
sentiment index is created as the percentage of positive (negative) words in 
the television news coverage of climate change each month, and serves as a 
proxy for optimism (pessimism) towards climate change. Our goal is to 
examine for an asymmetric effect of negative and positive sentiments, if any, 
on the risk profile and stock prices of clean energy firms in the US.
% Fifth Paragraph
%------------------------------------------------------------------------------

The climate risk and sentiment metrics, discussed in the previous paragraphs, 
are constructed at the monthly level using data between December 
2013 to August 2021. We then employ fixed-effect regressions to analyze how 
climate risk and sentiment measures--alongside the government's COVID-19 
policy, firm-specific, and macroeconomic variables--affect the idiosyncratic 
and systematic risks of 48 clean energy US firms. Our results show that
the volume of coverage, which serves as an indicator of aggregate climate 
risk, has a significant and positive effect on systematic risk.  
This provides evidence that the stock prices of clean energy firms account 
for climate risks, and as climate risk increases, so does systematic risk. 
Consequently, investors may demand higher returns to compensate for the 
elevated risk. Conversely, the volume of coverage has a negative effect on 
idiosyncratic risk. This implies that an increase in climate-related news 
decreases the volatility of returns for clean energy firms and makes it 
attractive to investors. 

When the climate risk is segregated to physical and transition risks, we find 
that climate crisis (an indicator for physical risks) and government \& human 
initiatives (which represent a form of transition risk) positively influences 
systematic risk. This, in turn, raises investors' expectation for stock 
returns from clean energy firms. However, neither physical nor transition 
risks have any significant effect on the firms' idiosyncratic risk. On the 
impact of sentiment indices, our results confirm the negativity bias found in 
existing literature. We observe that negative sentiments tend to decrease 
idiosyncratic risk and increase systematic risk, while positive sentiments 
have no significant impact. In addition, we conduct three robustness check to 
ensure that our results are reliable and not unduly affected by the presence 
of print media sentiment variables (CH Negative Climate Change News index 
\citep{Engle-etal-2020} and Media Climate Change Concerns 
\citep{Ardia-etal-2023}), climate policy uncertainty index 
\citep{Gavriilidis-2021}, and disruptions during the COVID-19 period. 

%-----------------------------------------------------------------------------
This article makes several contributions to the literature 
on climate finance. \emph{First}, this is the first study that utilizes 
television news coverage to create measures of climate risks and sentiment 
metrics. \emph{Second}, by analyzing the impact of metrics derived from 
television news on firms' risk profiles, it offers a new perspective on 
whether financial markets adjust asset prices for climate risks or if these 
risks are simply firm-specific traits that investors can eliminate through 
diversification. \emph{Third}, it establishes the impact of climate risks and 
sentiment variables on the risk profiles of clean energy firms in the US. 
\emph{Fourth}, it provides evidence that the information obtained from 
television news is distinct from those captured by economic policy 
uncertainty and print media sources, which have been prevalent in climate 
finance research.

The remainder of the paper is organized as follows. Section~\ref{sec:data} 
explains the data used in the analysis. Section~\ref{sec:EmpAnalysis} 
introduces the econometric model and discusses how the constructed metrics 
affect the systematic and idiosyncratic risks of clean energy firms. 
Section~\ref{sec:RobustChecks} presents a variety of robustness checks
and finally, Section~\ref{sec:conclusion} provides concluding remarks and 
points to directions for future research.

%------------------------------------------------------------------------------
\section{Data}\label{sec:data}
%------------------------------------------------------------------------------

In this section, we present the data that has been compiled 
from a variety of sources and utilized to conduct the study. 
Section~\ref{sec:DataClimate} uses television news data from the Global 
Dataset of Events, Language, and Tone (GDELT) database, and constructs six 
novel climate risk variables classified into three categories. 
Section~\ref{sec:DataFirmMacro} 
describes the firm-level and macroeconomic variables which are used as 
controls in the regressions. Finally, Section~\ref{sec:DataRisks} explains 
the construction of dependent variables i.e., systematic and idiosyncratic 
risks, using the three factor Fama-French model.

\subsection{Climate Risk Data}\label{sec:DataClimate}
%-----------------------------------------------------------------------------

We construct multiple climate risk measures from the television broadcast 
news data collected from the GDELT database. The GDELT covers news in more 
than 100 languages from print, broadcast, and web media beginning January 1, 
1979. We extract television broadcast news data from the GDELT’s Television 
Explorer, an interface that allows users to browse television news using 
keyword search (beginning July 2009) for more than 160 national, local 
US, and some international news stations. To sample relevant news data for 
our climate risk measures, we need to select (a) US news channels, and (b) 
relevant climate related keywords. 

With respect to choice of financial news stations, we choose Bloomberg, CNBC, 
and Fox Business. This is done for a couple of reasons. First, a typical 
audience of these US based channels comprise of 
active stock market participants for whom television news contributes to 
decision making. Second, news stations are constrained by airtime 
availability, so financial news channels are more likely to focus on 
important issues pertaining to the financial market. We extract news data for 
the three stations between December 2013 to August 2021. Our time frame is 
limited by data availability because Bloomberg 
does not provide television news data prior to December, 2013.

\begin{sloppypar}
Our selection of climate related keywords are guided by three conditions: 
underlying data, goals on variable construction, and caution against data 
mining (i.e., to avoid over fitting in in-sample prediction and poor 
performance in out-of-sample prediction) as suggested by 
\citep{Engle-etal-2020}. The underlying data are snippets from the Television 
Explorer, where a snippet is a 15-seconds block of airtime, 
excluding advertisements. Our goal is to construct climate risk measures, so 
our choice of keywords relate to climate change vocabulary such that the 
largest number of relevant snippets are selected. Such a dictionary based 
approach with some manual supervision, unlike machine learning methods, 
minimizes false positives and negatives \citep{Li-etal-2024}.
So, we judiciously select our keywords as opposed to entire vocabulary, such 
as in \citet{Engle-etal-2020}, because a vocabulary may include words (e.g., 
methane, nitrogen, and weather) which may not relate to climate change. With 
the above conditions in mind, we examined 26 climate change glossaries%
%------------------------
\footnote{Auburn University, BBC, Cambridge, Canadian Broadcasting 
Corporation (CBC), Center for Climate and Energy Solutions (C2ES), CNBC, 
Conservation in Changing Climate, DASolar, EDF, EMS Environmental, European 
Climate Adaptation, Platform (Climate ADAPT), European Environmental Agency 
(EEA), Global Greenhouse Warming, IPCC Special Report, MacMillan, National 
Geographic, New York Times, Statewide Integrated Flora and Fauna Teams 
(SWIFFT), The Guardian, The National Academy of Sciences, US Energy 
Information Administration (EIA), UK Climate Impacts Programme (UKCIP), 
United Nations Framework Convention on Climate Change (UNFCCC), University of 
California Davis, University of Miami, and Wikipedia.\label{fn1}} % 
%------------------------ 
and used the following keywords: 
\texttt{black carbon}, \texttt{cap and trade}, \texttt{carbon intensity}, 
\texttt{carbon budget}, \texttt{carbon emission}, \texttt{carbon footprint}, 
\texttt{carbon market}, \texttt{carbon tax}, \texttt{climate change}, 
\texttt{climate crisis}, \texttt{climate feedback}, \texttt{CO2}, 
\texttt{conference of the parties}, \texttt{COP 16}, \texttt{COP 21}, 
\texttt{emissions trading}, \texttt{global warming}, \texttt{greenhouse 
effect}, \texttt{greenhouse gases}, \texttt{intergovernmental panel on 
climate change}, \texttt{ipcc}, \texttt{Kyoto protocol}, \texttt{Montreal 
protocol}, \texttt{Paris agreement}, \texttt{renewable energy}, and 
\texttt{UNFCCC}. Based on these keywords, we obtain a total of 37,948 
snippets from the three news channels. For the purpose of analysis, we 
aggregate all snippets from the three channels for each month. 
\end{sloppypar}

The extracted snippets (of 15 seconds each) are climate change specific data 
and they relate to physical risk as well as transition risk, which is a 
leading concern for financial news channels. We employ textual analysis of 
snippets to define climate risk measures which are likely to affect decisions 
of clean energy stock market participants whenever discussions are on issues 
of climate change. To define the climate risk measures, we classify the 
content into three categories: \emph{volume of coverage}, \emph{type of 
coverage}, and the \emph{type of media sentiment} reflected in the language.  

\begin{figure*}[!t]
	\centerline{
		\mbox{\includegraphics[scale=0.42]{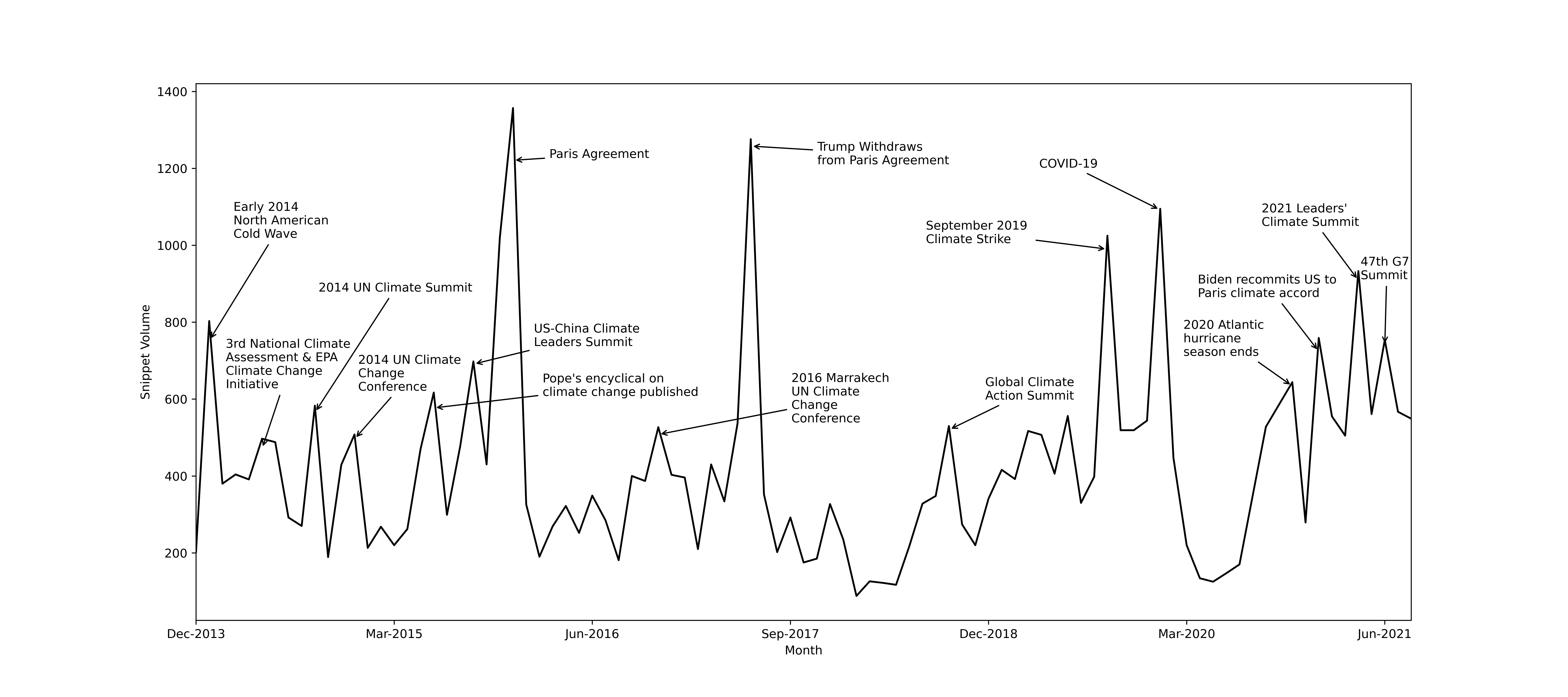}}
		%\mbox{\includegraphics[width=7.5in, height=3.25in]{Fig1-new.png}}
	}
	\caption{Volume of snippets per month with annotation of major events.} 
	\label{fig:VolCov}
\end{figure*}

The first measure i.e., volume of coverage of climate change ($VolCov$) is 
defined as the (logarithm) number of climate-change snippets in each month 
($t$). A plot of the number of snippets per month is presented in 
Figure~\ref{fig:VolCov} with annotation of major events. It is apparent from 
the figure that climate change coverage increases around major events such as 
international climate summit (e.g., September 2019 Climate Summit), large 
natural disasters (e.g., COVID-19), and important changes to climate 
regulation (e.g., Paris Agreement in April 2016). 

%---------------------------  Figure 1 ---------------------------------------

%------------------------------------------------------------------------------

To understand the second measure i.e., type of climate change coverage, we 
rely on our understanding of the coverage pattern and construct vocabulary 
lists for three main themes – \emph{climate crisis}, \emph{renewable energy}, 
and \emph{government \& human initiatives}. The \emph{climate crisis} theme, 
as the name suggests, focuses on crises due to changes in the physical 
climate (e.g., pollution, global warming, hurricanes, wildfire etc). In 
constrast, \emph{renewable energy} theme focuses on clean 
energy technologies (e.g., solar, wind, and other green energy). Lastly, the 
\emph{government \& human initiatives} theme capture governmental and 
international regulations, and climate 
adaption \& mitigation strategies (e.g., carbon tax, afforestation, 
biodiversity restoration, etc.). We construct the vocabulary list by 
scanning 26 relevant sources (see Footnote~\ref{fn1})
and assigning climate change-related 
term to each of the three themes. In total, we include 154, 152, and 132 
wildcard words  in the climate crisis, renewable energy, 
and government and human initiative vocabularies, respectively. 

%---------------------------  Figure 2 ---------------------------------------
\begin{figure*}[!t]
	\centerline{
		\mbox{\includegraphics[scale=0.5]{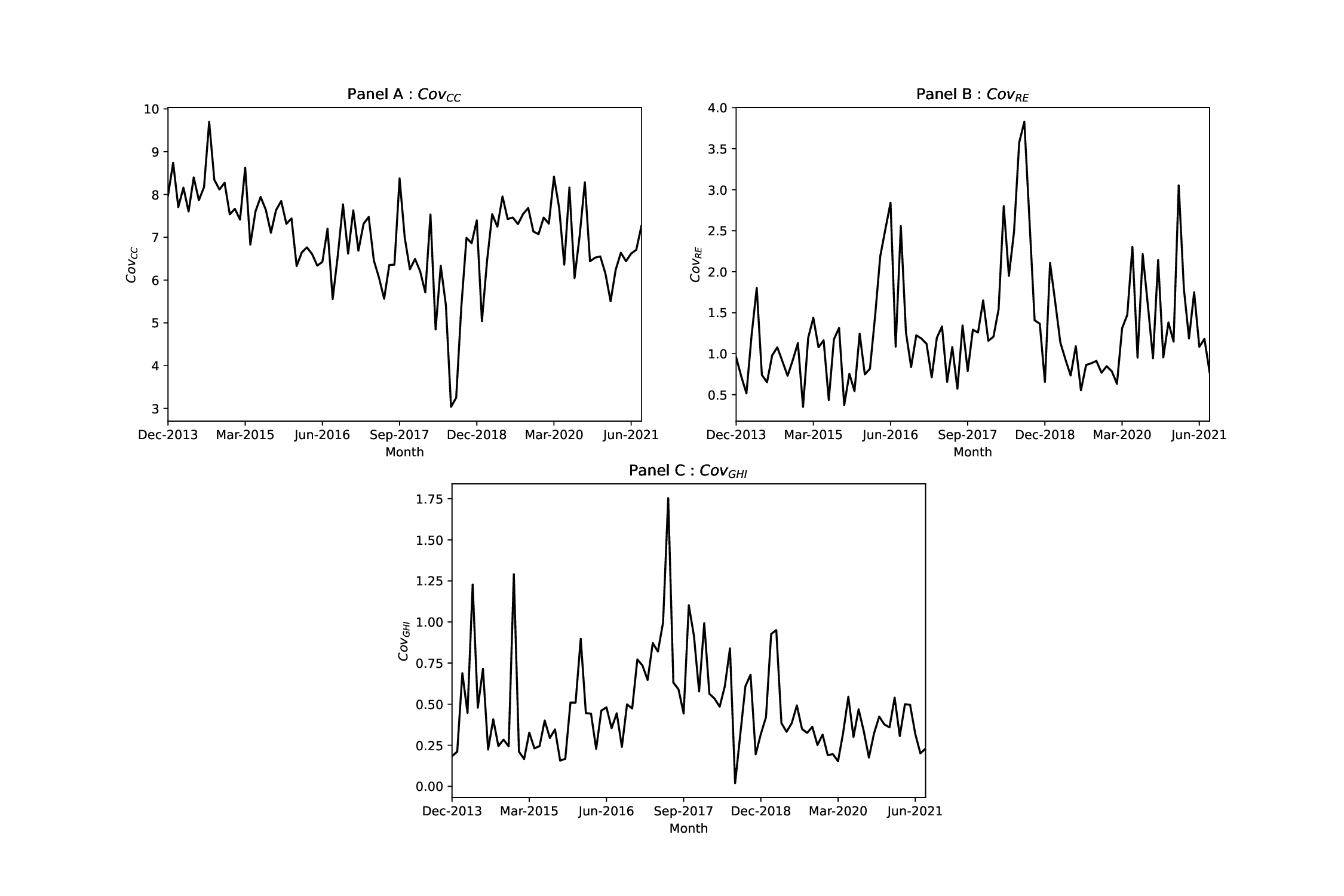}}
	}
	\caption{Type of coverage as percentage of total words.} 
	\label{fig:CovType}
\end{figure*}
%------------------------------------------------------------------------------

The coverage index for theme $i$ during month $t$ is 
calculated as: $Cov_{i,t}=N_{it}/(WC_{t})*100$,
where $N_{it}$ is the frequency of words related to theme $i$ during period 
$t$, and $WC_{t}$ is the total word count in all snippets during period 
$t$. Thus, $Cov_{i,t}$ represents the percentage of words in theme $i$ 
relative to the total number of words during time period $t$. In our study, 
$i$ can be climate crisis ($CC$), renewable energy ($RE$), or government and 
human initiatives ($GHI$). Our climate crises theme represents physical 
risk, whereas the renewable energy and government \& human initiative themes 
are part of transition risk. Figure~\ref{fig:CovType} displays a time series 
plot illustrating the monthly percentage of words associated with the three 
themes. It shows that $Cov_{CC}$ reached its lowest point near the Global 
Climate Action Summit in September 2018, while it peaked around the 2014 UN 
Climate Summit. In contrast, $Cov_{RE}$ was at its highest during the 2018 
Global Climate Action Summit, and $Cov_{GHI}$ saw its peak when President 
Donald Trump withdrew from the Paris Agreement. Analyzing these three themes 
together helps us understand how various types of climate news coverage 
impact the clean energy sector.
 
\begin{figure*}[!t]
	\centerline{
		\mbox{\includegraphics[scale=0.5]{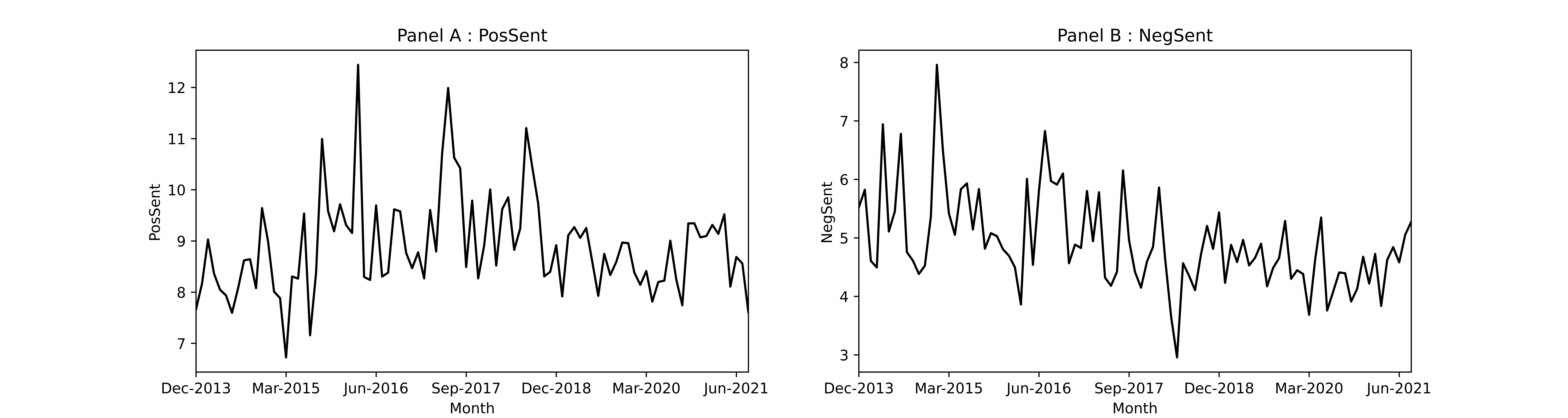}}
	}
	\caption{Type of sentiments as percentage of total words.} 
	\label{fig:Sent}
\end{figure*}

Finally, to capture news stations’ sentiment around climate change 
topics, we construct two sentiments variables in the spirit of 
\citet{Birz-Lott-2011} using the NRC Emotion Lexicon aka EmoLex 
\citep{Mohammad-Turney-2012}, a dictionary of two sentiments (negative or 
positive) and eight emotions (anger, anticipation, disgust, fear, joy, 
sadness, surprise, and trust). Specifically, the negative sentiment index is 
calculated as: $NegSent_{t} = \frac{NegW_{t}}{WC_{t}} \ast 100$, where 
$NegW_{t}$ depicts the frequency of negative words in month $t$ and $WC_{t}$ 
is as defined earlier. Thus, $NegSent_{t}$ shows pessimism in the language of 
television discourse around climate change for period $t$. 
Similarly, a positive sentiment index ($PosSent_{t}$) is defined 
as the frequency of positive words ($PosW_{t}$) divided by $WC_{t}$ and 
multiplied by 100. Figure~\ref{fig:Sent} presents a time series plot of  
positive and negative sentiments, and points to correlation with major events 
on climate change. For instance, positive sentiments peaked around the 
signing of Paris Agreement in April 2016, while negative (and positive) 
sentiments were at their highest (and lowest) around 2014 UN Climate Change 
Conference. Notably, negative sentiments reached their lowest point around 
the 2018 Global Climate Action Summit. Together, these sentiment variables 
allow us to analyze any potential asymmetric reactions within the clean 
energy stock market.

%---------------------------  Figure 3 ---------------------------------------
%------------------------------------------------------------------------------

A description of all the climate variables and basic data summary is 
presented in the first panel of Table~\ref{Table:DataSummary}. On 
average, there are $352 \approx \exp(5.863)$ snippets discussing climate 
change for the period under consideration. Regarding the thematic coverage, 
about 7 percent of the words, with a standard deviation of 1.05, pertain to 
climate crisis. The average percentage of words related to renewable 
energy (government and human initiatives) is 1.30 (0.48), with a standard 
deviation of 0.68 (0.29). With respect to media sentiment, we see the average 
number of words with positive sentiment (8.90) is higher than the average 
number of words with negative sentiment (4.92), although the standard 
deviation for the former is only somewhat higher than that of the latter.

%---------------------------  Table 3 ----------------------------------------
\begin{table}[!t]
\centering \footnotesize \setlength{\tabcolsep}{6pt} 
\setlength{\extrarowheight}{1.5pt}
\setlength\arrayrulewidth{1pt}
\caption{Data summary: The table presents the mean (Mean), median (Med), 
standard deviation (Std), and skewness (Skew) of the variables used in the 
fixed-effects regressions.}
\begin{tabular}{lp{8.5cm}r r r r }
\toprule
\textsc{variable}       & Description   & Mean   & Med  &  Std  &  Skew \\
\midrule       
& \texttt{Television coverage and sentiment variables}   &   &  &   & \\
\midrule
%------------------------------------------------------------------------------
$\ln VolCov$  & Logarithm of number of climate-change snippets      
              & 5.863  &  5.922   &  0.550   & $-0.089$  \\
$Cov_{CC}$    & Percent of words related to climate crisis 
			  & 7.002  &  7.135   &  1.051   & $-0.940$       \\
$Cov_{RE}$    & Percent of words related to renewable energy
			  & 1.299  &  1.158   &  0.685   & 1.494       \\
%------------------------------------------------------------------------------
$Cov_{GHI}$   & Percent of words related to government and human initiatives
			  & 0.478  &  0.421   &  0.287   & 1.608       \\
$PosSent$     & Percent of words with positive sentiment
			  & 8.896  &  8.767   &  0.951   & 1.077       \\
$NegSent$     & Percent of words with negative sentiment
			  & 4.923  &  4.727   &  0.800   & 0.939       \\
%------------------------------------------------------------------------------
\midrule
& \texttt{Firm-level variables}  &   &  &   & \\
\midrule
%------------------------------------------------------------------------------
ROA         & Return on Assets
		    & $-1.733$  &  4.300   &  23.75   & $-1.717$       \\
$\ln MktCap$& Logarithm of market capitalization 
	        &  13.741   & 13.774   &  2.772   & $-2.294$       \\
$Leverage$  & Total debt of a company divided by total capital
            &   2.503   &  3.211   &  1.794   & $-0.833$   \\
%------------------------------------------------------------------------------
$\ln StockVol$ 
            & Logarithm of volume of stocks ('000) exchanged in a day
		    &  9.287    &  9.111   &  1.727   &   0.359   \\
$\ln IntAsset$    
            & Logarithm of total value of a company's intangible assets 
            &  9.996    & 11.192   &  4.488   & $-1.221$   \\
$MBV$       & Market to book value
            & $-0.784$  &  1.010   & 37.616   & $-19.384$  \\
\midrule
%------------------------------------------------------------------------------
& \texttt{Macroeconomic controls}   &   &  &   & \\
\midrule
%------------------------------------------------------------------------------
$\ln PSE$   & Logarithmic returns of Arca Tech 100 index maintained by NYSE 
		    & $-0.442$  &  1.865   &  17.422   & $-8.423$   \\
$\ln MSCI$  & Logarithmic returns of MSCI world index (US \$) 
		    & $-0.248$  &  1.251   & 10.053    & $-6.997$   \\
$\ln OVX$   & Logarithm of returns in crude oil volatility index
		    & 0.353     & $-3.027$ & 23.165    & 1.459   \\
%------------------------------------------------------------------------------
$\ln EPU$   & Logarithm of economic policy uncertainty index 
		    &  5.179    &  5.162   &  0.390   & 0.131   \\
$\ln Covid \times PS$
            & Interaction of logarithm of global COVID death and policy 
            stringency (e.g., lockdown and other restrictions) 
		    &  9.933    &  0       & 20.382   & 1.582   \\
%------------------------------------------------------------------------------
\midrule
& \texttt{Systematic and Idiosyncratic Risks}   &   &  &   & \\
\midrule
%------------------------------------------------------------------------------
SysRisk     & Sytematic risk, calculated using the three factor Fama-French 
model (FF3) 
		    &  1.290  &  1.195   &  1.310   & 1.143   \\
IdRisk      & Idiosyncratic risk, calculated using FF3 
			&  2.346  &  1.194   &  1.672   & 4.277   \\
\bottomrule
\end{tabular}
\label{Table:DataSummary}
\end{table}
%------------------------------------------------------------------------------
 
We anticipate our coverage and media sentiment variables will have an effect 
on the clean energy sector in a certain way. Following 
\citet{Engle-etal-2020}, we assume that ``all news is bad news,'' implying 
that any increase in climate change coverage volume and negative 
sentiments entail higher climate risks. Regarding the type of coverage, 
climate crises can be viewed as disruptions to economic activity, thereby 
posing an operational risk to all firms. Thus, news coverage about climate 
crises may negatively influence the clean energy stocks. In addition, news 
about renewable energy and pro-environmental government and human initiatives 
typically represent positive news for the clean energy sector. However, news 
coverage about government and human initiatives could imply policy change, 
which could increase uncertainty. This increases transition risks and, 
ultimately, market risks. Lastly, we expect negative and positive media 
sentiments to have an asymmetric effect on clean energy stocks.

\subsection{Firm Level and Macroeconomic Data}\label{sec:DataFirmMacro}
%-----------------------------------------------------------------------------

While our primary goal is to analyze the effect of climate risk variables 
on the idiosyncratic and systematic risks (see Section~\ref{sec:DataRisks}) 
of 48 clean energy firms\footnote{Acuity Brands, Advanced Emissions 
Solutions, Advanced Energy Industries, Air Products and Chemicals, Ameresco, 
American Superconductor, Amtech Systems, Amyris, Badger Meter, CECO 
Environmental, Ceres Power Holdings, China Longyuan Power Group, China 
Everbright International, Cia Energetica De Minas Grais, Codexis, Comtec 
Solar Systems Group, Cree, Daqo New Energy Corporation, Enel Americas, Energy 
Recovery, EnerSys, ESCO Technologies, First Solar, Franklin Electric, 
Gentherm, Green Plains, Hexcel, Itron, JinkoSolar Holding, LSI Industries, 
MYR Group, Nextera Energy, ON Semiconductor, Orion Energy Systems, Ormat 
Technologies, Plug Power, Power Integrations, Pure Cycle, Quanta Services, 
Renesola, REX American Resources, Sunpower, Tesla, Universal Display, Veeco 
Instruments, Vicor Corporation.}, %%
we utilize several firm-level and 
macroeconomic variables as controls to limit the influence of confounding and 
extraneous variables, and thereby reduce omitted variable bias. Data for all 
the firm-level variables and the first three macroeconomic variables (PSE, 
MSCI, and OVX) are taken from Thomson Reuters; whereas for other variables 
they are curated from various sources and stated when we discuss the 
variable. The variables are described in Table~\ref{Table:DataSummary} along 
with the basic data summary.

With respect to firm-level, we have six variables as described in the second 
panel of Table~\ref{Table:DataSummary}. The variable return on assets is used 
as a proxy for a firm's profitability. Similarly, market capitalization 
(i.e., total value of a company's outstanding shares) is used to proxy 
market's perception 
of a company's total equity value. We also include financial leverage which 
is the use of debt (borrowed funds) to invest in assets. Leverage is often 
used as a measure of excessive risk taking and hence a higher leverage 
indicates a risky bet for potential investors. The volume of stocks exchanged 
in a day indicates a company's liquidity and higher volume is considered 
better for short-term trading. On the other hand, value of intangible assets 
add to a company's future worth and can be far more valuable than tangible 
assets. Lastly, MBV is the ratio of a company's book value to its market 
value and is utilized by investors as an indicator of market's perception of 
a particular stock's value.  

To cover the macroeconomic aspect, we include five variables as described in 
the third panel of Table~\ref{Table:DataSummary}. Arca Tech 100 (PSE) 
index, maintained by New York Stock Exchange, is a price weighted index 
composed of common stocks and ADRs of technology related companies listed on 
the US stock exchange. Returns to PSE is used as a proxy for average return 
from technology related firms. We also include the MSCI World index, a 
popular measure of the global stock market that tracks the performance of 
large and mid-cap companies across 23 developed countries. To measure energy 
market uncertainty, we utilize the crude oil volatility index (OVX) from the 
Chicago Board Option Exchange. In addition, we include the US economic policy 
uncertainty (EPU) index as a measure of uncertainties in economic policy. 
Data for EPU is taken from \citet{Baker-etal-2016}. Lastly, 
we include the interaction of global COVID-19 death (data source: 
\url{https://github.com/CSSEGISandData/COVID-19}) with policy stringency 
\citep[data taken from][]{Hale-etal-2021} to control for the negative impact 
of the COVID-19 period.

\subsection{Systematic and Idiosyncratic Risk} \label{sec:DataRisks}
%-----------------------------------------------------------------------------

In this section, we explain the construction of dependent variables i.e., 
the systematic and idiosyncratic (or non-systematic) risks from the three 
factor Fama-French model (FF3).  The FF3 is an extension of the Capital Asset 
Pricing Model (CAPM) and aims to explain stock returns of a company based on 
three factors: market risk, size premium, and value premium. Specifically, 
the equation for firm/stock $i$ at time $t$ is expressed as,
%---------------------
\begin{equation}
(R_{it} - r_{ft}) = \alpha_{it} + \beta_{ER} (R_{Mt} -  r_{ft}) + \beta_{SMB} 
SMB_{t} + \beta_{HML} HML_{it} + \varepsilon_{it}, 
\label{eq:FF3}
\end{equation}
%---------------------
where $R_{it}$ denotes the total return of stock $i$ at time $t$, $r_{ft}$ 
is the risk free rate of return at time $t$, and $R_{Mt}$ is the total market 
portfolio return at time $t$. The difference $(R_{it} - r_{ft})$ is the 
expected excess return of stock $i$ at time $t$. On the right hand side, the 
first factor $(R_{Mt} - r_{ft})$ is the excess return 
on the market portfolio index (i.e., the difference between daily market 
return proxied by the S\&P500 index and the risk free return) at time $t$. 
The second factor $SMB$ captures the size effect and is defined as the excess 
return of small-cap companies over big-cap companies. Lastly, the third 
factor $HML$ represents value premium and is defined as the spread in returns 
between companies with a high book-to-market ratio and companies with a low 
book-to-market ratio. 

The coefficient $\beta_{ER}$ represents sensitivity to market risk and is 
typically used as a measure of systematic risk; while the standard deviation 
of $\varepsilon_{it}$, denoted $\sigma_{\varepsilon}$, is used as an 
idiosyncratic risk. We adopt these definitions and estimate FF3 models on 
daily data for each trading month for the period December 2013 to August 
2021. Note that monthly data is the unit of analysis in our panel data 
models. The estimates $\hat{\beta}_{ER}$ and 
$\hat{\sigma}_{\varepsilon}$ obtained from the regression, as outlined in 
equation~\eqref{eq:FF3}, serve as the dependent variables in subsequent 
analysis.

%------------------------------------------------------------------------------
\section{Empirical Analysis} \label{sec:EmpAnalysis}
%------------------------------------------------------------------------------

With the dependent and independent variables (media sentiment, 
firm, and macroeconomic variables) available, we estimate the 
following fixed-effects regression \citep{Greene-2017},
%-----------------------
\begin{equation}
y_{it} = \sum_{j=1}^{J} x'_{ij,t} \beta_{j} + \alpha + \gamma_{i} + \xi_{it}, 
\end{equation}%
%-----------------------
where $y_{it}$ is either the estimated idiosyncratic risk 
($\hat{\sigma}_{\epsilon}$) or estimated systematic risk ($\hat{\beta}_{ER}$) 
for firm $i$ at month $t$, $x'_{it}$ 
is a vector of independent variables (comprising of climate risk, sentiment, 
firm-level, and macroeconomic variables) described in 
Table~\ref{Table:DataSummary}, $\alpha$ is a constant, $\gamma_{i}$ denotes 
firm level fixed-effect, and $\xi_{it}$ denotes the error term. The 
estimation results from the regression of idiosyncratic risk 
on the independent variables is presented in 
Table~\ref{Table:IdioRisk}, and those from the regression of systematic risk 
on the same covariates is displayed in Table~\ref{Table:SysRisk}.

We see from column M1 in Table~\ref{Table:IdioRisk} that the volume of 
climate change coverage $(VolCov)$ has a negative impact on idiosyncratic 
risk (or firm risk); a one unit change in $\ln VolCov$ decreases 
idiosyncratic risk by 12.42 percentage points. An increase in climate news 
coverage increases public and investors' awareness on climate issues 
\citep{Sampei-Ayogi-Usui-2009}. If investors 
perceive clean energy firms as valuable, they may be more willing to invest, 
resulting in better risk and debt management by a firm 
\citep{El-Ouadghiri-etal-2021}. Consequently, this 
will decrease the idiosyncratic risk associated with the firms. In contrast, 
$\ln VolCov$, as seen from the column M1 in Table~\ref{Table:SysRisk}, has a 
positive impact on systematic risk and a one unit change in $\ln VolCov$ 
increases systematic risk by 6.45 percentage 
points. This result follows from the fact that market risk tend to 
increase in light of higher climate change coverage. Overall, we find 
comprehensive support for the claim that volume of media coverage has a 
significant effect on the idiosyncratic and systematic risks of clean energy 
firms in the US.

%------------------------------- Table 2 -------------------------------------
\begin{table}[!t]
\centering \small \setlength{\tabcolsep}{4pt}
\setlength{\extrarowheight}{1.0pt}
\setlength\arrayrulewidth{1pt} \caption{Results (coefficient estimates and 
robust standard errors in parenthesis) from fixed-effects regression 
of idiosyncratic risk on television coverage, media sentiment, firm, and 
macroeconomic variables. $\ast \ast$ and $\ast$  denote significance 
at 1 and 5 percents, respectively}\vspace{10pt}
\begin{tabular}{llr rrr rrr}
\toprule
&& \textsc{M1} & \textsc{M2}& \textsc{M3} 
&\textsc{M4} & \textsc{M5} & \textsc{M6} \\
\midrule
%-----------------------------------------------------------------------------
Constant && $0.3469$  &$-0.6331$  & $-0.3997$ & 
$-0.3968$  & $-0.4207$ & $0.3117$ & \\
\rowcolor[gray]{0.92}
&& $(1.2746)$  & $(1.2274)$  &$(1.1801)$ & 
$(1.1892)$  & $(1.1816)$ & $(1.2027)$ &\\		
%-----------------------------------------------------------------------------
$\ln VolCov$ &&   $-0.1242^{**}$  & $..$  & $..$ &  
$..$  & $..$ & $..$ & \\
\rowcolor[gray]{0.92}
&&  $(0.0409)$  & $..$  & $..$ & 
$..$  & $..$ & $..$ &\\		
%-----------------------------------------------------------------------------
$Cov_{CC}$ &&  $..$  & $0.0200$  & $..$ &  
$..$  & $..$ & $..$ &\\
\rowcolor[gray]{0.92}
&& $..$  & $(0.0213)$  &$..$ & 
$..$  & $..$ & $..$ & \\
%-----------------------------------------------------------------------------
$Cov_{RE}$ &&  $..$  &$..$  & $-0.0170$ &  
$..$  & $..$ & $..$ & \\
\rowcolor[gray]{0.92}
&& $..$  & $..$  &$(0.0273)$ & 
$..$  & $..$ &  $..$ &\\
%-----------------------------------------------------------------------------
%-----------------------------------------------------------------------------
$Cov_{GHI}$ &&   $..$  & $..$  & $..$ &  
$-0.0162$  & $..$ & $..$ & \\
\rowcolor[gray]{0.92}
&&  $..$  & $..$  &$..$ & 
$(0.0641)$  & $..$ & $..$ & \\
%-----------------------------------------------------------------------------
$PosSent$ && $..$  &$..$  &$..$ &  
$..$  & $0.0020$ & $..$ & \\
\rowcolor[gray]{0.92}
&& $..$  &$..$  &$..$ & 
$..$  & $(0.0183)$ & $..$ & \\
%-----------------------------------------------------------------------------
$NegSent$ &&  $..$  &$..$  &$..$ & 
$..$  & $..$ & $-0.0859^{**}$ &  \\
\rowcolor[gray]{0.92}
&& $..$  &$..$  &$..$ & 
$..$  & $..$ & $(0.0294)$ &\\
%-----------------------------------------------------------------------------
%-----------------------------------------------------------------------------
$ROA$ && $-0.0019$  &$-0.0019$  & $-0.0019$ & 
$ -0.0019$  & $-0.0019$ &  $-0.0018$ &\\
\rowcolor[gray]{0.92}
&& $(0.0014)$  & $(0.0014)$  &$(0.0014)$ & 
$(0.0014)$  & $(0.0014)$ & $(0.0014)$ &\\
%-----------------------------------------------------------------------------
$\ln MktCap$ && $0.0381$  &$0.0384$  & $0.0383$ & 
$0.0382$  & $0.0382$& $0.0368$ &\\
\rowcolor[gray]{0.92}
&& $(0.0332)$  & $(0.0333)$  &$(0.0333)$ & 
$(0.0333)$  & $(0.0333)$& $(0.0331)$ &\\
%-----------------------------------------------------------------------------
$Leverage$ && $0.0185$  &$0.0186$  & $0.0185$ & 
$0.0183$  & $0.0183$ & $0.0179$ & \\
\rowcolor[gray]{0.92}
&& $(0.0183)$  & $(0.0184)$  &$(0.0184)$ & 
$(0.0183)$  & $(0.0184)$ & $(0.0182)$ &\\
%-----------------------------------------------------------------------------
$\ln StockVol$ && $0.2055^{**}$  &$0.2046^{**}$  & $0.2039^{**}$ & 
$0.2040^{**}$  & $0.2040^{**}$ & $0.2019^{**}$ & \\
\rowcolor[gray]{0.92}
&&  $(0.0622)$  & $(0.0617)$  &$(0.0616)$ & 
$(0.0616)$  & $(0.0616)$ & $(0.0623)$ &\\
%-----------------------------------------------------------------------------
$\ln IntAsset$ && $0.0105$  &$0.0103$  & $0.0103$ & 
$0.0103$  & $0.0103$ & $0.0106$ & \\
\rowcolor[gray]{0.92}
&& $(0.0147)$  & $(0.0149)$  &$(0.0149)$ & 
$(0.0149)$  & $0.0149$ & $(0.0147)$ &\\
%-----------------------------------------------------------------------------
$MBV$ && $0.0010^{**}$  &$0.0011^{**}$  & $0.0011^{**}$ & 
$0.0010^{**}$  & $0.0010^{**}$ & $0.0010^{**}$ &\\
\rowcolor[gray]{0.92}
&& $(0.0001)$  & $(0.0001)$  &$(0.0001)$ & 
$(0.0001)$  & $(0.0001)$ & $(0.0001)$ &\\
%-----------------------------------------------------------------------------
$\ln PSE$ && $0.0308^{**}$  &$0.0362^{**}$  & $0.0363^{**}$ & 
$0.0359^{**}$  & $0.0361^{**}$ & $0.0351^{**}$ & \\
\rowcolor[gray]{0.92}
&& $(0.0091)$  & $(0.0089)$  &$(0.0089)$ & 
$(0.0091)$  & $(0.0088)$ & $(0.0089)$ &\\
%-----------------------------------------------------------------------------
$\ln MSCI$ && $-0.0512^{**}$  &$-0.0603^{**}$  & 
$-0.0604^{**}$ & $-0.0597^{**}$  & $-0.0601^{**}$ & $-0.0591^{**}$ & \\
\rowcolor[gray]{0.92}
&& $(0.0162)$  & $(0.0161)$  &$(0.0161)$ & 
$(0.0166)$  & $(0.0158)$ & $(0.0161)$ &\\
%-----------------------------------------------------------------------------
$\ln OVX$ && $0.0054^{**}$  &$0.0044^{**}$  & $0.0045^{**}$ & 
$0.0046^{**}$  & $0.0046^{**}$ & $0.0045^{**}$ &\\
\rowcolor[gray]{0.92}
&& $(0.0012)$  & $(0.0012)$  &$(0.0012)$ & 
$(0.0011)$  & $(0.0012)$ & $(0.0012)$ &\\
%-----------------------------------------------------------------------------
$\ln EPU$ && $-0.0022$  &$0.0213$  & $0.0086$ & 
$0.0056$  & $0.0054$ & $-0.0427$ & \\
\rowcolor[gray]{0.92}
&& $(0.1092)$  & $(0.1102)$  &$(0.1098)$ & 
$(0.1089)$  & $(0.1090)$ & $(0.1097)$ &\\
%-----------------------------------------------------------------------------
$\ln Covid\times PS$ && $0.0152^{**}$  &$0.0148^{**}$  & 
$0.0150^{**}$ & $0.0149^{**}$  & $0.0149^{**}$ & $0.0144^{**}$ & \\
\rowcolor[gray]{0.92}
&& $(0.0022)$  & $(0.0022)$  &$(0.0022)$ & 
$(0.0022)$  & $(0.0022)$ & $(0.0022)$ &\\
\midrule
%-----------------------------------------------------------------------------
\textrm{R--squared} && $0.0867$ &$0.0845$  &$0.0843$  & $0.0843$ & $0.0843$  
& 0.0865& \\
%-----------------------------------------------------------------------------
\bottomrule
\end{tabular}
\label{Table:IdioRisk}
\end{table}
%-----------------------------------------------------------------------------

%------------------------------- Table 3 -------------------------------------
\begin{table}[!t]
\centering \small \setlength{\tabcolsep}{4pt}
\setlength{\extrarowheight}{1.0pt}
\setlength\arrayrulewidth{1pt} \caption{Results (coefficient estimates and 
robust standard errors in parenthesis) from fixed-effects regression 
of systematic risk on television coverage, media sentiment, firm, and 
macroeconomic variables. $\ast \ast$, $\ast$, and $\dagger$  denote 
significance at 1, 5, and 10 percents, respectively. }\vspace{10pt}
\begin{tabular}{llr rrr rrr}
\toprule
&& \textsc{M1} & \textsc{M2}& \textsc{M3} 
&\textsc{M4} & \textsc{M5} & \textsc{M6} \\
\midrule		
%-----------------------------------------------------------------------------
Constant && $-0.2643$  &$-0.4264$  & $0.1328$ & $ 
0.0728$  & $-0.1284$ & $-0.3650$ & \\
\rowcolor[gray]{0.92}
&& $(0.7266)$  & $(0.6744)$  &$(0.7040)$ & 
$(0.7216)$  & $(0.7707)$ & $(0.7316)$ &\\
%-----------------------------------------------------------------------------
$\ln VolCov$ &&  $0.0645^{**}$  & $..$  & $..$ &  
$..$  & $..$ & $..$ &  \\
\rowcolor[gray]{0.92}
        &&  $(0.0409)$  & $..$  & $..$ & 
$..$  & $..$ & $..$ &\\		
%-----------------------------------------------------------------------------
$Cov_{CC}$ &&  $..$  & $0.0483^{*}$  & $..$ &  $..$  & $..$ & $..$ &\\
\rowcolor[gray]{0.92}
      && $..$  & $(0.0197)$  &$..$ & $..$  & $..$ & $..$ & \\
%-----------------------------------------------------------------------------
$Cov_{RE}$ &&  $..$  &$..$  & $-0.0256$ &  $..$  & $..$ & $..$ & \\
\rowcolor[gray]{0.92}
      && $..$  & $..$  &$(0.0241)$ & $..$  & $..$ & $..$ & \\
%-----------------------------------------------------------------------------
%-----------------------------------------------------------------------------
$Cov_{GHI}$ &&   $..$  & $..$  & $..$ &  
$0.1166^{\dagger}$  & $..$ & $..$ & \\
\rowcolor[gray]{0.92}
      &&  $..$  & $..$  &$..$ & $(0.0618)$  & $..$ & $..$ & \\
%-----------------------------------------------------------------------------
$PosSent$ && $..$  &$..$  &$..$ & $..$  & $0.0301$ & $..$ & \\
\rowcolor[gray]{0.92}
      && $..$  &$..$  &$..$ & $..$  & $(0.0216)$ & $..$ & \\
%-----------------------------------------------------------------------------
$NegSent$ && $..$  &$..$  &$..$ & 
$..$  & $..$ & $0.0589^{**}$ &  \\
\rowcolor[gray]{0.92}
	  && $..$  &$..$  &$..$ & $..$  & $..$ & $(0.0205)$ &\\
%-----------------------------------------------------------------------------
%-----------------------------------------------------------------------------
$ROA$ 	&& $-0.0005$  &$-0.0004$  & $-0.0004$ & 
$-0.0004$  & $-0.0005$& $-0.0005$  & \\
\rowcolor[gray]{0.92}
	    && $(0.0007)$ & $(0.0007)$  &$(0.0007)$ & 
$(0.0007)$  & $(0.0007)$ & $(0.0007)$ &\\
%-----------------------------------------------------------------------------
$\ln MktCap$ && $-0.0358$  &$-0.0356$  & $-0.0358$ & 
$-0.0359$  & $-0.0363$ & $-0.0349$ & \\
\rowcolor[gray]{0.92}
&& $(0.0271)$  & $(0.0272)$  &$(0.0271)$ & 
$(0.0272)$  & $(0.0273)$ & $(0.0268)$ &\\
%-----------------------------------------------------------------------------
$Leverage$ && $-0.0106$  &$-0.0099$  & $-0.0102$ & $ 
-0.0105$  & $-0.0109$ & $-0.0102$ & \\
\rowcolor[gray]{0.92}
&& $(0.0156)$  & $(0.0157)$  &$(0.0157)$ & 
$(0.0156)$  & $(0.0158)$ & $(0.0157)$ &\\
%-----------------------------------------------------------------------------
$\ln StockVol$ && $0.1716^{**}$  &$0.1737^{**}$  & $0.1723^{**}$ & 
$ 0.1724^{*}$  & $0.1726^{*}$ & $0.1739^{**}$ & \\
\rowcolor[gray]{0.92}
&& $(0.0638)$  & $(0.0632)$  &$(0.0639)$ & 
$(0.0645)$  & $(0.0646)$ & $(0.0633)$ &\\
%-----------------------------------------------------------------------------
$\ln IntAsset$ && $-0.0082$  &$-0.0080$  & $-0.0080$ & 
$-0.0083$  & $-0.0082$ & $-0.0082$ & \\
\rowcolor[gray]{0.92}
&& $(0.0090)$  & $(0.0091)$  &$(0.0090)$ & 
$(0.0089)$  & $(0.0090)$ & $(0.0092)$ &\\
%-----------------------------------------------------------------------------
$MBV$ && $-0.0012^{**}$  &$-0.0011^{**}$  & $-0.0012^{**}$ & 
$-0.0012^{**}$  & $-0.0012^{**}$ & $-0.0012^{**}$ & \\
\rowcolor[gray]{0.92}
&& $(0.0001)$  & $(0.0001)$  &$(0.0001)$ & 
$(0.0001)$  & $(0.0001)$ & $(0.0001)$ &\\
%-----------------------------------------------------------------------------
$\ln PSE$ && $-0.0184^{**}$  &$-0.0207^{**}$  & $-0.0207^{**}$ 
& $-0.0201^{**}$  & $-0.0206^{**}$ & $-0.0206^{**}$ &\\
\rowcolor[gray]{0.92}
&& $(0.0061)$  & $(0.0066)$  &$(0.0066)$ & 
$(0.0067)$  & $(0.0065)$& $(0.0067)$ &\\
%-----------------------------------------------------------------------------
$\ln MSCI$ && $0.0344^{**}$  &$0.0384^{**}$  & $0.0384^{**}$ & 
$0.0370^{**}$  & $0.0376^{**}$ & $0.0384^{**}$ & \\
\rowcolor[gray]{0.92}
&& $(0.0101)$  & $(0.0109)$  &$(0.0109)$ & 
$(0.0112)$  & $(0.0108)$ & $(0.0110)$ &\\
%-----------------------------------------------------------------------------
$\ln OVX$ && $0.0006$  &$0.0006$  & $0.0009$ & $ 
0.0010$  & $0.0010$ & $0.0011$ & \\
\rowcolor[gray]{0.92}
&& $(0.0009)$  & $(0.0009)$  &$(0.0009)$ & 
$(0.0009)$  & $0.0009$ & $(0.0009)$ &\\
%-----------------------------------------------------------------------------
$\ln EPU$ && $0.0383$  &$0.0724$  & $0.0390$ & 
$ 0.0337$  & $0.0325$ & $0.0674$ & \\
\rowcolor[gray]{0.92}
&& $(0.0684)$  & $(0.0671)$  &$(0.0685)$ & 
$(0.0683)$  & $(0.0681)$ & $(0.0717)$ &\\
%-----------------------------------------------------------------------------
$\ln Covid \times PS$ && $-0.0017$  &$-0.0018$  & $-0.0015$ & 
$-0.0012$  & $-0.0014$ & $-0.0012$ & \\
\rowcolor[gray]{0.92}
&& $(0.0011)$ &$(0.0011)$  & $(0.0011)$  &$(0.0011)$ & 
$(0.0011)$  & $(0.0011)$ &\\
\midrule
%-----------------------------------------------------------------------------
R--\textrm{squared} && $0.0156$  &$0.0163$  & $0.0155$ & 
$0.0155$  & $0.0153$& $0.0160$ &\\
%-----------------------------------------------------------------------------
\bottomrule
\end{tabular}
\label{Table:SysRisk}
\end{table}
%-----------------------------------------------------------------------------

Next, we analyze the impact of different types of 
climate news coverage--climate crisis ($Cov_{CC}$), renewable energy 
($Cov_{RW}$), and government \& human initiatives ($Cov_{GHI}$)--on the 
risk profile of clean energy firms. The results for idiosyncratic and 
systematic risks are presented in columns M2 to M4 
of Table~\ref{Table:IdioRisk} and Table~\ref{Table:SysRisk}, respectively. We 
see that the type of coverage is not important for idiosyncratic risk as the 
coefficients for $Cov_{CC}$, $Cov_{RE}$, and $Cov_{GHI}$ are statistically 
insignificant at 5 percent level (the default significance level). In 
contrast, the systematic risk is positively associated with coverage of 
climate crisis ($Cov_{CC}$) and government and human initiatives 
($CC_{GHI}$), with the latter being significant only at the 10 percent 
significance level. $Cov_{CC}$, which represents physical risk, works 
through the channel of economic disruption caused by climate crises. For 
example, hurricanes can damage roads and impair industrial \& economic 
activities by disrupting transportation facilities. This will have an adverse 
impact on the operation of clean energy firms, leading to increase in total 
and systematic risks. Our results align with those of 
\citet{Balvers-etal-2017}, \citet{Bansal-etal-2019}, and 
\citet{Nagar-Schoenfeld-2021}, who also observe significant impacts of 
physical risk. On the other hand, the arrival of news 
related to government and human initiatives, which reflects transition risk, 
increases policy uncertainty, thereby increasing systematic risk. Our finding 
agrees with \citet{Faccini-etal-2023}, where they find that only transition 
risk through government intervention is priced in the US stocks.  To 
summarize, we find notable variation in how different types of climate 
coverages affect the risk profiles of clean energy firms.

The results presented in Tables~\ref{Table:IdioRisk} and 
\ref{Table:SysRisk} also reveal that investors exhibit an asymmetric reaction 
to positive and negative climate news. We observe that the coefficient for 
positive sentiment $(PosSent)$ is statistically insignificant, but that of 
negative sentiment $(NegSent)$ is statistically significant (even at 1 
percent significance level). Therefore, investors or participants in the 
clean energy stock market react strongly to negative sentiments, but not so 
much to positive sentiments. Such an asymmetric response to media sentiments, 
but constructed from news reports and articles, is consistent with the 
findings of \citet{Huang-etal-2014}, \citet{Heston-Sinha-2017}, 
\citet{Bajo-Raimondo-2017}, \citet{Huang-etal-2018}, and 
\citet{He-etal-2024}. Consequently, our results confirm the negativity bias 
documented in the climate finance literature.

Beyond the negativity bias, we find that $NegSent$ has a negative effect on 
idiosyncratic risk. While it is counter intuitive to think that an increase 
in negative news about climate change would reduce the idiosyncratic risk of 
clean energy firms, we may observe such scenarios through shifts in 
investments driven by ESG considerations. When climate risk increases, 
thereby increasing $NegSent$, it can prompt investors to reassess their 
portfolios and seek investments that align with sustainability goals 
\citep{El-Ouadghiri-etal-2021}. As a result, clean energy firms 
may see increased investor interest and capital inflows, reducing 
their idiosyncratic risk related to funding and liquidity 
issues. On the other hand, $NegSent$ has a positive effect on systematic risk 
since an increase in negative news increases the overall and systematic risks.

Besides the climate risk and media sentiment variables introduced in this 
paper, Tables~\ref{Table:IdioRisk} and \ref{Table:SysRisk} also report 
the coefficients of some common firm-level and macroeconomic variables 
typically found in this type of research. We see from 
Table~\ref{Table:IdioRisk} that $\ln StockVol$, $MBV$, $\ln PSE$, $\ln OVX$, 
and $\ln Covid \times PS$ positively affect idiosyncratic risk, while $\ln 
MSCI$ has a negative effect. These findings align with our intuition and 
existing literature. For instance, higher trading volume ($\ln StockVol$) 
generally enhances market liquidity, which tend to reduce idiosyncratic risk. 
Likewise, an increase in $MBV$ reflects elevated growth expectations, which 
can lead to greater price volatility and higher idiosyncratic risk. An 
increase in returns for technology firms 
(as represented by $\ln PSE$), energy market uncertainty (given by $\ln 
OVX$), and policy stringency during the COVID-19 period (given by the 
interaction term $\ln Covid \times PS$) all contribute positively to 
idiosyncratic risk as expected. Lastly, higher returns from the global stock 
market ($\ln MSCI$) are associated with a reduction in the idiosyncratic risk 
of clean energy firms.

Turning to the results for systematic risk shown in 
Table~\ref{Table:SysRisk}, we find that $\ln StockVo$l and 
$\ln MSCI$ have a positive effect, whereas $MBV$ and $\ln PSE$ have a 
negative effect on the systematic risk. Once again, these findings align with 
the explanations provided earlier and are consistent with established 
literature. For example, a higher $MBV$ may signal stable and predictable 
earnings, thereby decreasing a firm's exposure to systematic risk due to 
reduced sensitivity to economic fluctuations. The other variables, such as 
$\ln OVX$ and $\ln Covid \times PS$, which are significant for idiosyncratic 
risk, do not show statistical significance for systematic risk.

%------------------------------------------------------------------------------
\section{Robustness Checks} \label{sec:RobustChecks}
%------------------------------------------------------------------------------
In this section, we examine the robustness of our results in the presence of 
alternative modes of media sentiments, climate coverage variables that 
distinguishes between policy initiatives versus physical occurrences, and the 
disturbances during the COVID-19 period.

%-----------------------------------------------------------------------------
\subsection{Robustness with Print Media Sentiment 
Variable}\label{sec:RobustPrintMedia}

In Section~\ref{sec:EmpAnalysis}, we have shown that negative sentiment in 
television news coverage of climate change has a significant effect on both 
idiosyncratic and systematic risks of clean energy firms. Our finding  
resonates with existing research, which has documented the impact 
of sentiment from print media on firms' risk profiles. For instance, 
\citet{Huang-etal-2018} finds that media sentiment (positive and negative) 
extracted from news report is positively associated with firm's total and 
idiosyncratic risks. Similarly, \citet{Bask-etal-2024} find that negative 
media sentiment, derived from news articles published in Financial Times, is 
an important factor to explaining stock returns in various asset pricing 
models.

%------------------------------- Table 4 -------------------------------------
\begin{table}[!b]
\centering \small \setlength{\tabcolsep}{10pt}
\setlength{\extrarowheight}{1.0pt}
\setlength\arrayrulewidth{1pt} \caption{The coefficient estimates and 
robust standard errors (in parenthesis) from fixed-effect regression 
in the presence of CH Negative Index from \citet{Engle-etal-2020} and the 
MCCC Index from \citet{Ardia-etal-2023}. $\ast \ast$, $\ast$, and 
$\dagger$  denote significance at 1, 5, and 10 percents, respectively. 
}\vspace{10pt}
\begin{tabular}{llr rrr rr}
\toprule
&& \multicolumn{2}{c}{Idiosyncratic Risk} && \multicolumn{2}{c}{Systematic 
	Risk}\\
\cline{3-4} \cline{6-7}
&& $CHNeg$ & $MCCC$ && $CHNeg$ & $MCCC$ & \\
\midrule		
%-----------------------------------------------------------------------------
Constant    && $2.0007$  &$2.2393^{\dagger}$  && $-2.4774^{*}$  & $-1.7467$ & 
\\
\rowcolor[gray]{0.92}
&& $(1.3631)$ &$(1.3224)$  && $(1.2254)$  &$(1.2833)$ & \\
%-----------------------------------------------------------------------------
$NegSent$   && $-0.0569^{\dagger}$ & $-0.0675^{*}$  && $0.0646^{**}$  & 
$0.0521^{*}$ & \\
\rowcolor[gray]{0.92}
&& $(0.0305)$ &$(0.0302)$  && $(0.0217)$  &$(0.0208)$ & \\
%-----------------------------------------------------------------------------
$CHNeg$ && $-73.5957^{*}$ & $..$  && $283.9628^{**}$  & 
$..$ & \\
\rowcolor[gray]{0.92}
&& $(36.2728)$ & $..$  && $(47.1770)$  &$..$ & \\
%-----------------------------------------------------------------------------
$MCCC$ && $..$ & $-0.2189^{*}$  && $..$  & 
$-0.0844$ & \\
\rowcolor[gray]{0.92}
&& $..$ & $(0.0974)$  && $..$  & $(0.0773)$ & \\
%-----------------------------------------------------------------------------
$ROA$ && $-0.0021$ & $-0.0021$  && $-0.0010$  & 
$-0.0006$ & \\
\rowcolor[gray]{0.92}
&& $(0.0018)$ & $(0.0017)$  && $(0.0010)$  & $(0.0009)$ & \\
%-----------------------------------------------------------------------------
$\ln MktCap$ && $ 0.0093$ & $ 0.0068$  && $-0.0478$  & 
$-0.0444$ & \\
\rowcolor[gray]{0.92}
&& $(0.0507)$ & $(0.0504)$  && $(0.0403)$  & $(0.0397)$ & \\
%-----------------------------------------------------------------------------
$Leverage$ && $ 0.0036$ & $ 0.0069$  && $-0.0176$  & 
$-0.0272$ & \\
\rowcolor[gray]{0.92}
&& $(0.0226)$ & $(0.0212)$  && $(0.0256)$  & $(0.0240)$ & \\
%-----------------------------------------------------------------------------
$\ln StockVol$ && $ 0.0978$ & $ 0.0773$  && $ 0.1339$  & 
$0.1464^{\dagger}$ & \\
\rowcolor[gray]{0.92}
&& $(0.0798)$ & $(0.0771)$  && $(0.0799)$  & $(0.0859)$ & \\
%-----------------------------------------------------------------------------
$\ln IntAsset$ && $ 0.0018$ & $ 0.0057$  && $-0.0016$  & 
$-0.0032$ & \\
\rowcolor[gray]{0.92}
&& $(0.0096)$ & $(0.0116)$  && $(0.0096)$  & $(0.0111)$ & \\
%-----------------------------------------------------------------------------
$MBV$ && $ 0.0017^{**}$ & $ 0.0016^{**}$  && $-0.0012^{**}$  & 
$-0.0013^{**}$ & \\
\rowcolor[gray]{0.92}
&& $(0.0001)$ & $(0.0001)$  && $(0.0001)$  & $(0.0001)$ & \\
%-----------------------------------------------------------------------------
$\ln PSE$ && $-0.0306^{\dagger}$ & $-0.0349^{*}$  && $-0.0454^{**}$  & 
$-0.0592^{**}$ & \\
\rowcolor[gray]{0.92}
&& $(0.0174)$ & $(0.0165)$  && $(0.0141)$  & $(0.0133)$ & \\
%-----------------------------------------------------------------------------
$\ln MSCI$ && $ 0.0409$ & $ 0.0443$  && $0.0838$  & 
$0.1004^{**}$ & \\
\rowcolor[gray]{0.92}
&& $(0.0274)$ & $(0.0270)$  && $(0.0160)$  & $(0.0166)$ & \\
%-----------------------------------------------------------------------------
$\ln OVX$ && $ 0.0032$ & $0.0030$  && $0.0023$  & 
$0.0036^{\dagger}$ & \\
\rowcolor[gray]{0.92}
&& $(0.0021)$ & $(0.0021)$  && $(0.0017)$  & $(0.0018)$ & \\
%-----------------------------------------------------------------------------
$\ln EPU$ && $-0.1088$ & $-0.0511$  && $ 0.6052^{**}$  & 
$ 0.4640^{**}$ & \\
\rowcolor[gray]{0.92}
&& $(0.1330)$ & $(0.1329)$  && $(0.1355)$  & $(0.1380 )$ & \\
%-----------------------------------------------------------------------------
\midrule
R--\textrm{squared} && $0.0095$ &$0.096$  && $0.0449$  & $0.0275$ & \\
%-----------------------------------------------------------------------------
\bottomrule
\end{tabular}
\label{Table:PrintMedia}
\end{table}
%-----------------------------------------------------------------------------

However, the impact of sentiments derived from television news coverage is 
distinct to those resulting from print media coverage. To substantiate this, 
we regress the systematic and idiosyncratic risks on $NegSent$ and 
measures of climate change news via print media channel, while controlling 
for firm and macroeconomic variables. In the first model, 
we include the CH Negative Climate Change News index ($CHNeg$) from 
\citet{Engle-etal-2020}. The $CHNeg$ index measures the proportion of total 
articles with negative tones in the complete collection of articles from 
various news sources, which are filtered based on the keyword ``climate 
change''. Whereas in the second model, we incorporate the Media Climate 
Change Concerns ($MCCC$) index from \citet{Ardia-etal-2023}.  
The MCCC index measures the concern in media about climate change 
using climate change news in 10 newspapers and 2 news wires published in the 
US. The data for $CHNeg$ and $MCCC$ are in months and available until May 
2018 and June 2018, respectively. Accordingly, the estimation results for the 
robustness of print media variables belong to a smaller time period, i.e., 
Dec, 2013--May, 2018 for models that include $CHNeg$ and Dec, 
2013--Jun, 2018 for models that incorporate $MCCC$.

The regression results for the print media variables $CHNeg$ and 
$MCCC$ are shown in Table~\ref{Table:PrintMedia} and they reinforce the 
findings from Tables~\ref{Table:IdioRisk} and 
\ref{Table:SysRisk}. When regressing idiosyncratic risk on $CHNeg$ ($MCCC$) 
and other control variables, the coefficient for $NegSent$ is significant at 
the 10\% (5\%) significance level. However, when systematic risk is regressed 
on $CHNeg$ or $MCCC$ and other control variables, $NegSent$ is significant at 
5\% or lower significance level. Overall, the significance of the television 
climate change sentiment variable, $NegSent$, remains robust even when 
accounting for the print media sentiment indices. This suggests that 
television media offers additional insights to investors beyond 
what is provided by print media and news wires, thereby improving their 
decision-making process for a better portfolio.

%------------------------------------------------------------------------------
\subsection{Robustness with Climate Policy Uncertainty}\label{sec:RobustCPU}

In the previous section, we analyzed the impact of two print media variables 
on firms risk profile, but both $CHNeg$ and $MCCC$ do not distinguish between 
coverage of climate change-related physical events and climate change policy 
initiatives. However, it is crucial to separate the effects of news coverage 
on policy initiatives, as governments worldwide are enacting various policies 
to combat climate change. While the policies differ in scope and execution 
from one country to another, they impact both the idiosyncratic and 
systematic risks of clean energy companies, thereby affecting their stock 
returns.

To achieve the aforementioned objective, we incorporate the climate policy 
uncertainty ($CPU$) index from \citet{Gavriilidis-2021} as a covariate in our 
regression models. The $CPU$ is a media-based policy uncertainty index, 
constructed by searching climate change related keywords (e.g., global 
warming, climate change) along with terms related to policy and uncertainty 
(e.g., regulation, White House, EPA, policy) in eight major U.S. newspapers:  
Boston Globe, Chicago Tribune, Los Angeles Times, Miami Herald, New York 
Times, Tampa Bay Times, USA Today and the Wall Street Journal. 
Table~\ref{Table:Robust-CPU} presents the empirical results, with columns 
(M1)-(M6) summarizing findings for idiosyncratic risk and columns (M7)-(M12) 
for systematic risk. With idiosyncratic as the dependent variable, we observe 
that both negative sentiment from television news coverage ($NegSent$) and 
coverage volume ($\ln VolCov$) are statistically 
significant at 1\% significance level, with minimal change in their 
coefficients. For the regressions with systematic risk, the coefficient for 
$\ln VolCov$ increases but that of $NegSent$ 
shows a minor decrease. Both variables remain significant at 5\% significance 
level. Thus, the influence of coverage volume and negative sentiment on the 
risk profiles of clean energy firms are robust to the inclusion of economic 
policy uncertainty index in the model.

%-----------------------------  Table 5 --------------------------------------
\afterpage{ %\clearpage
\begin{landscape}
\begin{table}%[!htbp]
%\begin{sidewaystable}
\centering \scriptsize \setlength{\tabcolsep}{4pt} 
\setlength{\extrarowheight}{0.5pt}
\setlength\arrayrulewidth{1pt} \caption{The table presents 
the coefficient estimates and robust standard errors (in parenthesis) from 
the regression of idiosyncratic and systematic risks on all the covariates in 
the presence of climate policy uncertainty (CPU) variable from 
\citet{Gavriilidis-2021}. $\ast \ast$ and $\ast$  denote 
significance at 1 and 5 percents, respectively.
}
%-----------------------------------------------------------------------------------------
\begin{tabular}{l rr rrr rr|r rrr rr }
\toprule
& \multicolumn{6}{c}{Idiosyncratic Risk} && \multicolumn{6}{c}{Systematic 
Risk}  \\
\cmidrule{2-8} \cmidrule{9-14}
&  M1  &  M2  &  M3  &  M4   &  M5   &  M6 &&
  M7  &  M8  &  M9  &  M10  &  M11  &  M12  \\
\midrule  
Constant  
&  $ 0.3441$  & $-0.7034$  & $-0.4367$ & $-0.4315$ & $-0.4335$ & $ 0.3122$   
&& $-0.2620$  & $-0.3882$  & $ 0.1700$ & $ 0.1206$ & $-0.1098$ & $-0.3344$
\\
\rowcolor[gray]{0.92}
& $(1.2795)$& $(1.2633)$ & $(1.2000)$& $(1.2059)$ &$(1.1893)$  & $(1.2457)$  
&& $(0.7241)$& $(0.6859)$ & $(0.7077)$& $(0.7223)$ &$(0.7681)$ & $(0.7522)$\\
%-----------------------------------------------------------------------------
$\ln VolCov$
&   $-0.1360^{**}$ & $ ..$    & $..$      & $..$      & $..$  & $ ..$ 
&&  $ 0.0741^{*}$  & $..$     & $..$      & $..$      & $..$  & $..$ 
\\
\rowcolor[gray]{0.92}
&  $(0.0410)$  & $..$     & $..$      & $..$       &$..$     & $..$
&& $(0.0326)$  & $..$     & $..$      & $..$       &$..$     & $..$
\\
%-----------------------------------------------------------------------------
$Cov_{CC}$
&  $..$      &$-0.1360$     & $..$      & $..$       & $..$   & $ ..$
&&  $..$     &$ 0.0471^{*}$ & $..$      & $..$       & $..$   & $..$ 
\\
\rowcolor[gray]{0.92}
& $..$       & $(0.0410)$   & $..$      & $..$       &$..$     & $..$      
&& $..$      & $(0.0203)$   & $..$      & $..$       &$..$     & $..$ 
\\
%-----------------------------------------------------------------------------
$Cov_{RE}$
&  $..$      & $..$         & $-0.0180$ & $..$       & $..$  & $..$   
&& $..$      & $..$         & $-0.0247$ & $..$       & $..$  & $..$   
\\
\rowcolor[gray]{0.92}
& $..$      & $..$          & $(0.0277)$& $..$       &$..$   & $..$     
&& $..$     & $..$          & $(0.0244)$& $..$       &$..$   & $..$     
\\
%-----------------------------------------------------------------------------
$Cov_{GHI}$
&  $..$      & $..$     & $..$       & $-0.0252$      & $..$  & $ ..$     
&& $..$      & $..$     & $..$       & $ 0.1289^{*}$  & $..$  & $..$      
\\
\rowcolor[gray]{0.92}
& $..$      &$..$      & $..$       & $(0.0610)$ &$..$   & $..$       
&& $..$     &$..$      & $..$       & $(0.0591)$ &$..$   & $..$     
\\
%-----------------------------------------------------------------------------
$PosSent$
&  $..$      & $..$     & $..$       & $..$       & $-0.0008$  & $ ..$     
&&  $..$     & $..$     & $..$       & $..$       & $ 0.0342$  & $..$      
\\
\rowcolor[gray]{0.92}
&  $..$      &$..$     & $..$       & $..$       &$(0.0196)$   & $..$       
&& $..$      &$..$     & $..$       & $..$       &$(0.0233)$   & $..$     
\\
%-----------------------------------------------------------------------------
$NegSent$
&  $..$      & $..$     & $..$       & $..$       &  $..$    & 
$-0.0859^{**}$ 
&&  $..$      & $..$     & $..$       & $..$       & $..$  & 
$0.0570^{*}$  
\\
\rowcolor[gray]{0.92}
& $..$       &$..$      & $..$       & $..$       &$..$    &$(0.0314)$   
&& $..$       &$..$      & $..$       & $..$       &$..$ &$(0.0226)$ 
\\
%-----------------------------------------------------------------------------
$CPU$
& $0.0926$    & $0.0541$ & $0.0464$   & $ 0.0484$  & $0.0454$  & $-0.0003$ 
&& $-0.0747$  & $-0.0294$& $-0.0467$  & $-0.0666$  & $-0.0663$ & $-0.0187$  
\\
\rowcolor[gray]{0.92}
& $(0.0636)$ &$(0.0662)$& $(0.0645)$ & $(0.0617)$ &$(0.0671)$  &$(0.0684)$   
&& $(0.0629)$ &$(0.0599)$& $(0.0587)$ & $(0.0560)$ &$(0.0632)$ &$(0.0625)$   
\\
%-----------------------------------------------------------------------------
$ROA$
& $-0.0019$  & $-0.0018$& $-0.0019$  & $-0.0019$  & $-0.0019$  & $-0.0018$ 
&& $-0.0005$ & $-0.0004$& $-0.0005$  & $-0.0004$  & $-0.0005$  & $-0.0005$  
\\
\rowcolor[gray]{0.92}
& $(0.0014)$  &$(0.0014)$& $(0.0014)$ & $(0.0014)$ &$(0.0014)$ &$(0.0014)$   
&& $(0.0007)$ &$(0.0007)$& $(0.0007)$ & $(0.0007)$ &$(0.0007)$ &$(0.0007)$   
\\
%-----------------------------------------------------------------------------
$\ln MktCap$
& $ 0.0383$  & $ 0.0385$& $ 0.0384$  & $ 0.0384$  & $ 0.0384$  & $ 0.0368$ 
&& $-0.0360$ & $-0.0357$& $-0.0359$  & $-0.0361$  & $-0.0366$  & $-0.0350$  
\\
\rowcolor[gray]{0.92}
& $(0.0333)$ &$(0.0334)$& $(0.0333)$ & $(0.0333)$ &$(0.0334)$  &$(0.0330)$    
&&$(0.0270)$ &$(0.0272)$& $(0.0271)$ & $(0.0271)$ &$(0.0273)$  &$(0.0268)$   
\\
%-----------------------------------------------------------------------------
$Leverage$
& $ 0.0187$  & $ 0.0187$& $ 0.0186$  & $ 0.0184$  & $ 0.0184$  & $ 0.0179$ 
&& $-0.0107$ & $-0.0099$& $-0.0103$  & $-0.0106$  & $-0.0110$  & $-0.0102$  
\\
\rowcolor[gray]{0.92}
& $(0.0183)$ &$(0.0184)$& $(0.0184)$ & $(0.0183)$ &$(0.0184)$  &$(0.0182)$   
&&$(0.0156)$ &$(0.0157)$& $(0.0157)$ & $(0.0155)$ &$(0.0158)$  &$(0.0157)$  
\\
%-----------------------------------------------------------------------------
$\ln StockVol$
& $ 0.2043^{**}$  & $ 0.2039^{**}$& $ 0.2033^{**}$  & $ 
0.2033^{**}$  & $ 0.2034^{**}$   & $ 0.2019^{**}$ 
&& $ 0.1726^{**}$  & $ 0.1741^{**}$& $ 0.1729^{**}$  & $ 
0.1734^{**}$  & $ 0.1736^{**}$ & $ 0.1741^{**}$ 
\\
\rowcolor[gray]{0.92}
& $(0.0621)$ &$(0.0614)$& $(0.0612)$ & $(0.0613)$ &$(0.0612)$  &$(0.0620)$    
&&$(0.0634)$ &$(0.0631)$& $(0.0636)$ & $(0.0643)$ &$(0.0644)$  &$(0.0632)$   
\\
%-----------------------------------------------------------------------------
$\ln IntAsset$
& $ 0.0102$  & $ 0.0101$& $ 0.0102$  & $ 0.0102$  & $ 0.0101$  & $ 0.0106$ 
&& $-0.0079$ & $-0.0079$& $-0.0078$  & $-0.0081$  & $-0.0079$  & $-0.0082$ 
\\
\rowcolor[gray]{0.92}
& $(0.0146)$ &$(0.0148)$& $(0.0148)$ & $(0.0148)$ &$(0.0148)$ &$(0.0147)$     
&&$(0.0091)$ &$(0.0091)$& $(0.0091)$ & $(0.0090)$ &$(0.0091)$ &$(0.0093)$    
\\
%-----------------------------------------------------------------------------
$MBV$
& $ 0.0010^{**}$  & $ 0.0011^{**}$& $ 0.0010^{**}$  & $ 
0.0010^{**}$  & $ 0.0010^{**}$   & $ 0.0010^{**}$ 
&& $-0.0011^{**}$  & $-0.0011^{**}$& $-0.0011^{**}$  & 
$-0.0011^{**}$  & $-0.0012^{**}$ & $-0.0012^{**}$ 
\\
\rowcolor[gray]{0.92}
& $(0.0001)$ &$(0.0001)$& $(0.0001)$ & $(0.0001)$ &$(0.0001)$  &$(0.0001)$   
&&$(0.0001)$ &$(0.0001)$& $(0.0001)$ & $(0.0001)$ &$(0.0001)$  &$(0.0001)$  
\\
%-----------------------------------------------------------------------------
$\ln PSE$
& $ 0.0287^{**}$  & $ 0.0354^{**}$& $ 0.0356^{**}$  & $ 
0.0350^{**}$  & $ 0.0353^{**}$   & $ 0.0351^{**}$ 
&& $-0.0168^{**}$  & $-0.0202^{**}$& $-0.0200^{**}$  & 
$-0.0189^{**}$  & $-0.0194^{**}$ & $-0.0203^{**}$ 
\\
\rowcolor[gray]{0.92}
& $(0.0085)$ &$(0.0085)$& $(0.0085)$ & $(0.0088)$ &$(0.0083)$  &$(0.0085)$ 
&&$(0.0062)$ &$(0.0066)$& $(0.0066)$ & $(0.0067)$ &$(0.0066)$  &$(0.0067)$  
\\
%-----------------------------------------------------------------------------
$\ln MSCI$
& $-0.0480^{**}$  & $-0.0589^{**}$& $-0.0593^{**}$  & 
$-0.0583^{**}$  & $-0.0588^{**}$   & $-0.0591^{**}$ 
&& $ 0.0319^{**}$  & $ 0.0377^{**}$& $ 0.0372^{**}$  & $ 
0.0351^{**}$  & $ 0.0357^{**}$ & $ 0.0379^{**}$ 
\\
\rowcolor[gray]{0.92}
& $(0.0154)$ &$(0.0154)$& $(0.0155)$ & $(0.0160)$ &$(0.0151)$  &$(0.0154)$    
&& $(0.0102)$ &$(0.0110)$& $(0.0110)$ & $(0.0112)$ &$(0.0109)$ &$(0.0110)$   
\\
%-----------------------------------------------------------------------------
$\ln OVX$
& $ 0.0056^{**}$  & $ 0.0045^{**}$& $ 0.0046^{**}$  & $ 
0.0046^{**}$  & $ 0.0047^{**}$   & $ 0.0045^{**}$  
&& $ 0.0004$  & $ 0.0006$& $ 0.0008$  & $ 0.0010$  & $ 0.0009$ & $ 0.0010$  
\\
\rowcolor[gray]{0.92}
& $(0.0012)$ &$(0.0012)$& $(0.0012)$ & $(0.0011)$ &$(0.0011)$  &$(0.0011)$   
&&$(0.0008)$ &$(0.0009)$& $(0.0009)$ & $(0.0009)$ &$(0.0009)$  &$(0.0009)$    
\\
%-----------------------------------------------------------------------------
$\ln EPU$
& $-0.0737$  & $-0.0182$& $-0.0266$  & $-0.0313$  & $-0.0291$  & $-0.0425$  
&&$ 0.0960$  & $ 0.0939$& $ 0.0745$  & $ 0.0845$  & $ 0.0829$  & $ 0.0806$  
\\
\rowcolor[gray]{0.92}
& $(0.0988)$ &$(0.0986)$& $(0.0987)$ & $(0.0996)$ &$(0.0994)$  &$(0.0987)$   
&&$(0.0806)$ &$(0.0763)$& $(0.0776)$ & $(0.0799)$ &$(0.0813)$  &$(0.0783)$   
\\
%-----------------------------------------------------------------------------
$\ln Covid \times PS$
& $ 0.0151^{**}$  & $ 0.0148^{**}$& $ 0.0150^{**}$  & $0.0148^{**}$  & $ 
0.0149^{**}$   & $ 0.0144^{**}$ 
&& $-0.0017$  & $-0.0017$& $-0.0015$  & $-0.0011$  & $-0.0013$  & $-0.0012$ 
\\
\rowcolor[gray]{0.92}
& $(0.0022)$ &$(0.0022)$& $(0.0022)$ & $(0.0022)$ &$(0.0022)$  &$(0.0022)$  
&& $(0.0011)$ &$(0.0011)$& $(0.0011)$ & $(0.0011)$ &$(0.0011)$ &$(0.0011)$   
\\ 
\midrule 
%-----------------------------------------------------------------------------
R--\textrm{squared} 
& $0.0871$ &$0.0846$& $0.0845$ & $0.0844$ &$0.0844$  &$0.0865$   
&& $0.0159$ &$0.0164$& $0.0151$ & $0.0158$ &$0.0156$ &$0.0161$ \\
\bottomrule
\end{tabular}
%-----------------------------------------------------------------------------
\label{Table:Robust-CPU}
%\end{sidewaystable}
\end{table}
\end{landscape}
}
%-----------------------------------------------------------------------------

%-----------------------------------------------------------------------------
\subsection{Robustness Check for the COVID-19 Period}

%-----------------------------  Table 6 --------------------------------------
\afterpage{
\begin{landscape}
\begin{table}%[!htbp]
%\begin{sidewaystable}
\centering \scriptsize \setlength{\tabcolsep}{4pt} 
\setlength{\extrarowheight}{0.5pt}
\setlength\arrayrulewidth{1pt} \caption{COVID-19 Phase. The table presents 
the coefficient estimates and robust standard errors (in parenthesis) from 
the regression of idiosyncratic and systematic risks on climate risk 
measures, firm and macroeconomic variables using data for the period: 
January, 2020--August, 2021. $\ast \ast$, $\ast$, and $\dagger$  denote 
significance at 1, 5, and 10 percents, respectively.
}
%-----------------------------------------------------------------------------------------
\begin{tabular}{l rr rrr rr|r rrr rr }
\toprule
& \multicolumn{6}{c}{Idiosyncratic Risk} && \multicolumn{6}{c}{Systematic 
	Risk}  \\
\cmidrule{2-8} \cmidrule{9-14}
&  M1  &  M2  &  M3  &  M4   &  M5   &  M6 &&
M7  &  M8  &  M9  &  M10  &  M11  &  M12  \\
\midrule  
Constant  
&  $-2.2028$  & $-8.4209^{**}$  & $-8.4126^{**}$ & $-8.6733^{**}$ 
&  $-10.5728^{**}$ & $-5.5474^{*}$   
&& $ 0.3689$  & $-0.5954$  & $-0.4603$ & $-0.7744$ & $-2.8642$ & $-0.3070$
\\
\rowcolor[gray]{0.92}
& $(3.8409)$ & $(2.1867)$ & $(2.1926)$& $(2.1113)$ &$(2.4508)$  & $(2.2156)$  
&& $(2.3335)$& $(1.4157)$ & $(1.4327)$& $(1.4864)$ &$(2.0259)$ & $(1.4712)$\\
%-----------------------------------------------------------------------------
$\ln VolCov$
&   $-0.3929^{*}$ & $ ..$    & $..$      & $..$      & $..$  & $ ..$ 
&&  $-0.0612$  & $..$     & $..$      & $..$      & $..$  & $..$ 
\\
\rowcolor[gray]{0.92}
&  $(0.1616)$  & $..$     & $..$      & $..$       &$..$     & $..$
&& $(0.0877)$  & $..$     & $..$      & $..$       &$..$     & $..$
\\
%-----------------------------------------------------------------------------
$Cov_{CC}$
&  $..$      &$-0.0017$     & $..$      & $..$       & $..$   & $ ..$
&&  $..$     &$-0.2544^{**}$ & $..$      & $..$       & $..$   & $..$ 
\\
\rowcolor[gray]{0.92}
& $..$       & $(0.0794)$   & $..$      & $..$       &$..$     & $..$      
&& $..$      & $(0.0605)$   & $..$      & $..$       &$..$     & $..$ 
\\
%-----------------------------------------------------------------------------
$Cov_{RE}$
&  $..$      & $..$         & $ 0.0157$      & $..$       & $..$  & $..$   
&& $..$      & $..$         & $ 0.2616^{**}$ & $..$       & $..$  & $..$   
\\
\rowcolor[gray]{0.92}
& $..$      & $..$          & $(0.0975)$& $..$       &$..$   & $..$     
&& $..$     & $..$          & $(0.0613)$& $..$       &$..$   & $..$     
\\
%-----------------------------------------------------------------------------
$Cov_{GHI}$
&  $..$      & $..$     & $..$       & $ 0.6572$      & $..$  & $ ..$     
&& $..$      & $..$     & $..$       & $ 0.4566$  & $..$  & $..$      
\\
\rowcolor[gray]{0.92}
& $..$      &$..$      & $..$       & $(0.4824)$ &$..$   & $..$       
&& $..$     &$..$      & $..$       & $(0.3176)$ &$..$   & $..$     
\\
%-----------------------------------------------------------------------------
$PosSent$
&  $..$      & $..$     & $..$       & $..$       & $ 0.2079$  & $ 
..$     
&&  $..$     & $..$     & $..$       & $..$       & $ 0.2188^{**}$  & 
$..$      
\\
\rowcolor[gray]{0.92}
&  $..$      &$..$     & $..$       & $..$       &$(0.1261)$   & $..$       
&& $..$      &$..$     & $..$       & $..$       &$(0.0801)$   & $..$     
\\
%-----------------------------------------------------------------------------
$NegSent$
&  $..$      & $..$     & $..$       & $..$       &  $..$    & 
$-0.3652^{**}$ 
&&  $..$      & $..$     & $..$       & $..$       & $..$  & 
$-0.0371$  
\\
\rowcolor[gray]{0.92}
&  $..$       &$..$      & $..$       & $..$       &$..$    &$(0.1300)$   
&& $..$       &$..$      & $..$       & $..$       &$..$    &$(0.0995)$ 
\\
%-----------------------------------------------------------------------------
$ROA$
& $-0.0048$  & $-0.0051$& $-0.0051$  & $-0.0051$  & $-0.0052$  & $-0.0051$ 
&& $-0.0017$ & $-0.0019$& $-0.0017$  & $-0.0017$  & $-0.0019$  & $-0.0017$  
\\
\rowcolor[gray]{0.92}
& $(0.0043)$  &$(0.0042)$& $(0.0041)$ & $(0.0041)$ &$(0.0042)$ &$(0.0042)$   
&& $(0.0021)$ &$(0.0023)$& $(0.0021)$ & $(0.0022)$ &$(0.0022)$ &$(0.0021)$   
\\
%-----------------------------------------------------------------------------
$\ln MktCap$
& $ 0.0700$  & $ 0.0786$& $ 0.0786$  & $ 0.0804$  & $ 0.0820$  & $ 0.0816$ 
&& $0.0329$  & $ 0.0386$& $ 0.0345$  & $ 0.0354$  & $ 0.0378$  & $ 0.0345$  
\\
\rowcolor[gray]{0.92}
& $(0.0716)$ &$(0.0721)$& $(0.0726)$ & $(0.0720)$ &$(0.0708)$  &$(0.0711)$    
&&$(0.0524)$ &$(0.0511)$& $(0.0505)$ & $(0.0521)$ &$(0.0493)$  &$(0.0526)$   
\\
%-----------------------------------------------------------------------------
$Leverage$
& $-0.0120$  & $-0.0256$& $-0.0256$  & $-0.0247$  & $-0.0286$  & $-0.0284$ 
&& $ 0.0160$ & $ 0.0104$& $ 0.0136$  & $ 0.0144$  & $ 0.0106$  & $ 0.0136$  
\\
\rowcolor[gray]{0.92}
& $(0.0269)$ &$(0.0258)$& $(0.0259)$ & $(0.0260)$ &$(0.0262)$  &$(0.0253)$   
&&$(0.0309)$ &$(0.0291)$& $(0.0307)$ & $(0.0297)$ &$(0.0290)$  &$(0.0303)$  
\\
%-----------------------------------------------------------------------------
$\ln StockVol$
& $ 0.4658^{**}$  & $ 0.4785^{**}$& $ 0.4781^{**}$  & 
$ 0.4711^{**}$  & $ 0.4707^{**}$   & $ 0.4796^{**}$ 
&& $ 0.2338^{**}$  & $ 0.2277^{**}$& $ 0.2286^{**}$  & $ 
0.2306^{**}$  & $ 0.2276^{**}$ & $ 0.2359^{**}$ 
\\
\rowcolor[gray]{0.92}
& $(0.1293)$ &$(0.1132)$& $(0.1145)$ & $(0.1169)$ &$(0.1105)$  &$(0.1146)$    
&&$(0.585)$ &$(0.0513)$& $(0.0573)$ & $(0.0565)$ &$(0.0521)$  &$(0.0561)$   
\\
%-----------------------------------------------------------------------------
$\ln IntAsset$
& $ 0.0135$  & $ 0.0153$ & $ 0.0152$  & $ 0.0144$  & $ 0.0169$  & $ 0.0187$ 
&& $-0.0259$  & $-0.0245$& $-0.0274$  & 
$-0.0262$  & $-0.0240$  & $-0.0252$ 
\\
\rowcolor[gray]{0.92}
& $(0.0395)$ &$(0.0383)$& $(0.0387)$ & $(0.0379)$ &$(0.0375)$ &$(0.0387)$     
&&$(0.0220)$ &$(0.0203)$& $(0.0217)$ & $(0.0215)$ &$(0.0207)$ &$(0.0218)$    
\\
%-----------------------------------------------------------------------------
$MBV$
& $ 0.1159^{**}$  & $ 0.1120^{**}$& $ 0.1121^{**}$  & 
$ 0.1119^{**}$  & $ 0.1104^{**}$   & $ 0.1117^{**}$ 
&& $0.0248$  & $ 0.0211$& $ 0.0240$ & $ 0.0241$  & $ 0.0224$ & $ 0.0241$ 
\\
\rowcolor[gray]{0.92}
& $(0.0107)$ &$(0.0127)$& $(0.0128)$ & $(0.0127)$ &$(0.0137)$  &$(0.0129)$   
&&$(0.0197)$ &$(0.0175)$& $(0.0198)$ & $(0.0193)$ &$(0.0177)$  &$(0.0192)$  
\\
%-----------------------------------------------------------------------------
$\ln PSE$
& $ 0.0324^{*}$  & $ 0.0676^{**}$& $ 0.0674^{**}$  
& $ 0.0672^{**}$  & $ 0.0765^{**}$   & $ 0.0638^{**}$ 
&& $0.0363^{**}$  & $ 0.0424^{**}$& $ 0.0390^{**}$  
&  $0.0415^{**}$  & $ 0.0511^{**}$ & $ 0.0414^{**}$ 
\\
\rowcolor[gray]{0.92}
& $(0.0157)$ &$(0.0168)$& $(0.0163)$ & $(0.0168)$ &$(0.0148)$  &$(0.0175)$ 
&&$(0.0134)$ &$(0.0104)$& $(0.0105)$ & $(0.0105)$ &$(0.0106)$  &$(0.0106)$  
\\
%-----------------------------------------------------------------------------
$\ln MSCI$
&  $-0.0578^{\dagger}$  & $-0.1223^{**}$   & $-0.1221^{**}$  
&  $-0.1229^{**}$  & $-0.1411^{**}$   & $-0.1172^{**}$ 
&& $-0.0635^{**}$  & $-0.0783^{**}$   & $-0.0708^{**}$  
&  $-0.0740^{**}$  & $-0.0934^{**}$ & $-0.0731^{**}$ 
\\
\rowcolor[gray]{0.92}
& $(0.0288)$ &$(0.0309)$& $(0.0307)$ & $(0.0313)$ &$(0.0271)$  &$(0.0322)$    
&& $(0.0234)$ &$(0.0180)$& $(0.0180)$ & $(0.0179)$ &$(0.0186)$ &$(0.0182)$   
\\
%-----------------------------------------------------------------------------
$\ln OVX$
& $ 0.0108^{**}$  & $ 0.0056^{*}$& $ 0.0056^{*}$  
& $ 0.0066^{**}$  & $ 0.0039^{*}$     & $ 0.0033^{\dagger}$  
&& $-0.0062^{**}$  & $-0.0048^{**}$& $-0.0068^{**}$  & $-0.0063^{**}$  
& $-0.0088^{**}$  & $-0.0073^{**}$  
\\
\rowcolor[gray]{0.92}
& $(0.0023)$ &$(0.0024)$& $(0.0021)$ & $(0.0022)$ &$(0.0019)$  &$(0.0020)$   
&&$(0.0018)$ &$(0.0016)$& $(0.0015)$ & $(0.0017)$ &$(0.0015)$  &$(0.0016)$    
\\
%-----------------------------------------------------------------------------
$\ln EPU$
& $ 0.2305$  & $ 0.8725^{**}$& $ 0.8675^{**}$  & $ 0.8969^{**}$  
& $ 0.9583^{**}$  & $ 0.6399^{**}$  
&&$-0.2533$  & $ 0.1845$& $-0.1993$  & $-0.1352$  & $-0.0610$  
& $-0.1771$  
\\
\rowcolor[gray]{0.92}
& $(0.3445)$ &$(0.1916)$& $(0.1893)$ & $(0.1774)$ &$(0.1869)$  &$(0.1805)$   
&&$(0.2184)$ &$(0.1997)$& $(0.1541)$ & $(0.1609)$ &$(0.1714)$  &$(0.1678)$   
\\
%-----------------------------------------------------------------------------
$\ln Covid \times PS$
& $ 0.0116^{**}$  & $ 0.0134^{**}$& $ 0.0131^{**}$  & $ 0.0119^{**}$  
& $ 0.0110^{**}$  & $ 0.0124^{**}$  
&&$ 0.0048^{**}$  & $ 0.0022$& $0.0011$  & $0.0040^{**}$  & $ 0.0026$  
& $ 0.0050^{**}$  
\\
\rowcolor[gray]{0.92}
& $(0.0029)$ &$(0.0031)$& $(0.0031)$ & $(0.0029)$ &$(0.0032)$  &$(0.0031)$   
&&$(0.0014)$ &$(0.0014)$& $(0.0015)$ & $(0.0015)$ &$(0.0016)$  &$(0.0013)$   
\\
\midrule 
%-----------------------------------------------------------------------------
R--\textrm{squared} 
& $0.1767$ &$0.1646$& $0.1646$ & $0.1659$ &$0.1687$  &$0.1720$   
&& $0.0504$ &$0.0724$& $0.0693$ & $0.0512$ &$0.0604$ &$0.0499$ \\
\bottomrule
\end{tabular}
%-----------------------------------------------------------------------------
\label{Table:CovidPhase}
%\end{sidewaystable}
\end{table}
\end{landscape}
}
%----------------------------------

The COVID-19 pandemic caused widespread disruptions in the global 
economy and affected the financial markets, including the renewable energy 
sector. For instance, \citet{Roy-etal-2022} 
find that the clean energy sector under performed during COVID-19 period. 
Consequently, we aim to investigate whether the climate risk and sentiment 
measures had any different effect on the idiosyncratic and systematic risks 
during the COVID-19 pandemic. To do this, we re-estimate our regression 
models from Table~\ref{Table:IdioRisk} and Table~\ref{Table:SysRisk} for the 
COVID-19 period: January, 2020 to August, 2021.

The results for the COVID-19 period are reported in 
Table~\ref{Table:CovidPhase}, where the first six columns (M1)-(M6) display 
the regression results of idiosyncratic risk while the remaining six columns 
(M7)-(M12) report the regression results of systematic risk. A comparison of 
the coefficients for climate risk measures in Table~\ref{Table:CovidPhase} 
with those from the full sample, in Tables~\ref{Table:IdioRisk} and 
\ref{Table:SysRisk}, reveals notable differences. Beginning with the volume 
of coverage, we find that for idiosyncratic risk the coefficient for $\ln 
VolCov$ increases in magnitude but is now significant at 5\% significance 
level. Whereas for systematic risk, the coefficient for $\ln VolCov$ changes 
from being positive and significant to being statistically insignificant.

Turning attention to type of coverage variables, we find that for 
idiosyncratic risk the coefficient for $Cov_{CC}$ remains statistically 
insignificant; whereas for systematic risk it changes from being positive to 
negative, and is now significant at 1\% level. So, during the pandemic, 
climate crises does not impact volatility of returns, but does reduce the 
expected stock return for clean energy firms. The variable $Cov_{RE}$ shows 
an insignificant effect on idiosyncratic risk in the full sample and COVID-19 
subsample, while for systematic risk the impact changes from being 
insignificant to positive and significant during the pandemic. 
The positive effect of $Cov_{RE}$ on systematic risk in the clean energy 
sector can be linked to higher uncertainty during the COVID-19 period, 
affecting the macroeconomic conditions. In contrast, the coefficient for 
$Cov_{GHI}$ becomes statistically insignificant for both idiosyncratic and 
systematic risks. This may indicate that transition risk was not that 
important during this period, as government efforts were targeted to contain 
the pandemic and vaccine development.

Lastly, for media sentiments, we find that the negative impact of $NegSent$ 
on idiosyncratic risk is significantly stronger in the COVID-19 subsample 
as compared to the full sample. For systematic risk, the effect of $NegSent$ 
changes from being positive and significant to insignificant during the 
pandemic. Interestingly, the positive sentiment variable $PosSent$, which has 
remained insignificant in our entire analysis, positively impacts the 
systematic risk in the COVID-19 subsample. In summary, 
during the COVID-19 period, volume of coverage and negative sentiment are 
crucial to understanding idiosyncratic risk; whereas the type of coverage and 
positive sentiment appears to be more important for explaining systematic 
risk of clean energy firms.

%------------------------------------------------------------------------------
\section{Conclusion}\label{sec:conclusion}
%------------------------------------------------------------------------------

This paper pioneers the use of television news coverage data to develop 
climate risk and sentiment measures, and utilizes these measures to assess 
the risk profiles of US clean energy firms. We focus on clean energy firms 
due to their critical role in combating climate change. Using monthly data 
from December 2013 to August 2021, we derive our climate risk and sentiment 
metrics from the television news coverage by Bloomberg, CNBC, and Fox 
Business; sourced from the Global Dataset of Events, Language, and Tone 
(GDELT) database. We then employ fixed-effects regression to explore how 
these metrics, along with government COVID-19 policies, firm-specific 
factors, and macroeconomic variables, influence the idiosyncratic and 
systematic risks of clean energy firms. Our results indicate that volume of 
coverage, which is a proxy for aggregate risk \citep{Engle-etal-2020}, 
positively affects systematic risk, which in turn will raise expected return. 
Moreover, the volume of coverage negatively affects idiosyncratic risk, 
suggesting that an increase in climate-related news decreases the volatility 
of returns for clean energy firms and makes it attractive to investors.

Next, following \citet{Faccini-etal-2023} we categorize aggregate risk into 
two types: physical risks related to climate-induced weather events and 
transition risks tied to policy changes regarding renewable energy and 
government \& human initiatives. Our analysis 
reveals that both physical risks and transition risks from government \& 
human initiatives have a positive impact on systematic risk. Consequently, 
this raises investors’ expectations for returns from clean energy firms. 
However, neither physical risks nor transition risks significantly affect the 
firms’ idiosyncratic risk. Our findings also support the negativity bias 
identified in existing research, showing that negative sentiments tend to 
decrease idiosyncratic risk and increase systematic risk, while positive 
sentiments have no notable effect. We conduct several robustness checks and 
find that innovations from television news coverage differ, and do not 
confound, from those arising 
from print media sources and climate policy uncertainty index. However, 
analysis during the COVID-19 period reveals some deviations from the results 
observed in the full sample.

Television news coverage in the GDELT database offers a goldmine of data for 
developing climate risk metrics and integrating them into asset pricing 
models, and thereby enriching the climate finance literature. With this 
extensive data, a range of research questions can be explored. Here, we 
highlight three key areas for investigation. First, television news coverage 
can be utilized to create more detailed measures of physical and transition 
risks \citep{Venturini-2022, Faccini-etal-2023}, 
and analyze their effect on firms risk profiles and stock returns. Second, 
new insights from television coverage can be employed to design a dynamic 
hedging portfolio, extending the approach of \citet{Engle-etal-2020}. Third, 
building on the work of \citet{Huynh-Xia-2021}, one can investigate whether 
climate risk measures derived from television news coverage are priced in the 
bond and options markets.  

%------------------------------------------------------------------------------
\section*{CRediT authorship contribution statement}\label{sec:CRediT}
%------------------------------------------------------------------------------
Wasim Ahmad: Conceptualization, Methodology, Resources, and Editing. Mohammad 
Arshad Rahman: Writing \& Editing. Suruchi Shrimali: Data 
Curation, Methodology, Analysis. Preeti Roy: Data Curation, Resources, and 
Analysis.

%------------------------------------------------------------------------------
\section*{Declaration of competing interest}
%------------------------------------------------------------------------------
None.

%--------------------------------- Bibliography ------------------------------
\clearpage \pagebreak %\nocite{*}
%\pdfbookmark[1]{References}{unnumbered} % To make References as a bookmark in
%pdf
%\section*{References}
\singlespacing
\bibliography{BibClimateRisk}
\bibliographystyle{jasa}

\end{document}